\documentclass[12pt]{article}
\usepackage{amsmath,amssymb}
\usepackage{bm}
\usepackage{color}
\usepackage{graphicx}
\usepackage{cite}
\usepackage{mathtools}
\usepackage{slashed}
\usepackage{braket}
\usepackage{breqn}

\usepackage{multirow}

\begin{document}

\title{
\begin{flushright}
\ \\*[-80pt]
\begin{minipage}{0.25\linewidth}
\normalsize
EPHOU-23-009\\*[50pt]
\end{minipage}
\end{flushright}
{\Large \bf
Zero-modes in magnetized  $T^6/\mathbb{Z}_N$ orbifold models 
through  $Sp(6,\mathbb{Z})$ modular symmetry
\\*[20pt]}}

\author{
Shota Kikuchi$^{1}$,
Tatsuo Kobayashi,$^{1}$
Kaito Nasu$^{1}$, \\
Shohei Takada$^{1}$, 
and
~Hikaru Uchida$^{1,2}$
\\*[20pt]
\centerline{
\begin{minipage}{\linewidth}
\begin{center}
$^{1}${\it \normalsize
Department of Physics, Hokkaido University, Sapporo 060-0810, Japan} \\*[5pt]
$^{2}${\it \normalsize
Institute for the Advancement of Graduate Education, Hokkaido University, Sapporo 060-0817, Japan} \\*[5pt]
\end{center}
\end{minipage}}
\\*[50pt]}

\date{
\centerline{\small \bf Abstract}
\begin{minipage}{0.9\linewidth}
\medskip
\medskip
\small
We study of fermion zero-modes on magnetized $T^6/\mathbb{Z}_N$ orbifolds. In particular, we focus on non-factorizable orbifolds, i.e. $T^6/\mathbb{Z}_7$ and $T^6/\mathbb{Z}_{12}$ corresponding to
$SU(7)$ and $E_6$ Lie lattices respectively.
The number of degenerated zero-modes corresponds to the generation number of low energy effective theory in four dimensional space-time. We find that three-generation models preserving 4D $\mathcal{N}=1$ supersymmetry can be realized by magnetized $T^6/\mathbb{Z}_{12}$, but not by $T^6/\mathbb{Z}_7$. We use $Sp(6,\mathbb{Z})$ modular transformation for the analyses. 
\end{minipage}
}

\begin{titlepage}
\maketitle
\thispagestyle{empty}
\end{titlepage}

\newpage

%
\tableofcontents
\clearpage
\section{Introduction}
\label{Intro}

Higher dimensional theory such as superstring theory is interesting as a candidate for unified theory of particle physics.
When we start with higher dimensional theory, we need compactification of extra dimensions.
In particular, compactifications leading to four-dimensional (4D) chiral theory are important, 
because the standard model is a chiral theory.

Inspired by superstring theory, we start with six-dimensional (6D) compact space.
One of the simplest compactifications is the toroidal compactification $T^6$.
However, that leads to 4D non-chiral theory.
One way to derive a 4D chiral theory is orbifolding $T^6/\mathbb{Z}_N$ \cite{Dixon:1985jw,Dixon:1986jc}.
4D supersymmetry (SUSY) must be broken to $\mathcal{N}=1$ or $0$ to realize a 4D chiral theory.
The $\mathbb{Z}_N$ twists to preserve 4D $\mathcal{N}=1$ SUSY were classified \cite{Dixon:1985jw,Dixon:1986jc}.
In addition, six-dimensional lattices with those $\mathbb{Z}_N$ twist symmetries were studied in Refs.~ 
\cite{Markushevich:1986za,Ibanez:1987pj,Katsuki:1989bf,Kobayashi:1991rp,Lust:2005dy,Lust:2006zg}.

Another way to lead to a 4D chiral theory is introduction of magnetic fluxes in compact space 
\cite{Bachas:1995ik,Berkooz:1996km,Blumenhagen:2000wh,Angelantonj:2000hi}.
The degeneracy number of zero-modes, which corresponds to the generation of 4D massless chiral fermions, 
is determined by the size of magnetic fluxes.
Yukawa couplings in 4D low-energy effective field theory are computed by overlap integrals of zero-mode wave functions \cite{Cremades:2004wa,Antoniadis:2009bg}.
They can lead to suppressed Yukawa couplings as well as ${\cal O}(1)$ of couplings depending on moduli values.

One can combine the above geometrical background and gauge background and study the orbifold compactification with 
magnetic flux background \cite{Abe:2008fi,Abe:2013bca}.
Adjoint matter fields can be projected out in magnetized orbifold models, and 
that corresponds to stabilization of Wilson line moduli, i.e. open string moduli in intersecting D-brane models on orbifolds 
\cite{Blumenhagen:2005tn}, 
which are T-dual to magnetized D-brane models on orbifolds.
Magnetized orbifold models have richer flavor structure.
Three-generation models can be derived by various setups on the $T^2/\mathbb{Z}_N$ orbifold with magnetic flux 
\cite{Abe:2008sx,Abe:2015yva}.
Furthermore, realization of quark and lepton mass matrices were studied 
\cite{Abe:2012fj,Abe:2014vza,Fujimoto:2016zjs,Buchmuller:2017vho,Buchmuller:2017vut,Kikuchi:2021yog,Hoshiya:2022qvr}.

So far, the six dimensional space, which can be factorizable to three two-dimensional spaces, was 
mainly studied, although some non-factorizable $T^4/\mathbb{Z}_N$ orbifolds were studied \cite{Kikuchi:2022lfv,Kikuchi:2022psj}.
Our purpose is to study non-factorizable cases.
Here, we study $T^6/\mathbb{Z}_N$ orbifold models with magnetic fluxes, whose $T^4$ or $T^6$ parts are non-factorizable.
We examine their zero-mode numbers.
In particular, we show three-generation models.
Such studies were done in magnetized $T^2/\mathbb{Z}_N$ orbifold models by several methods 
\cite{Abe:2008fi,Abe:2013bca,Sakamoto:2020pev,Kobayashi:2022tti,Imai:2022bke}.
Among them, one way to analyze zero-mode numbers in magnetized $T^2/\mathbb{Z}_N$ orbifold models is to 
use the $SL(2,\mathbb{Z})$ modular symmetry of wave functions on $T^2$ \cite{Kobayashi:2017dyu}.
(See also Ref.~\cite{Kikuchi:2022psj}.)
We extend such analysis to $T^6/\mathbb{Z}_N$ orbifolds as well as  $T^4/\mathbb{Z}_N$ orbifolds.
Higher dimensional compact spaces such as $T^6$ have several moduli, and have larger $Sp(2g,\mathbb{Z})$ 
symplectic modular symmetries.
(See for mathematical reviews, e.g. Refs.~\cite{Siegel,Mumford:1983}.)
These large $Sp(2g,\mathbb{Z})$ symplectic modular symmetries appear in string compacitication.
(See e.g. Refs.~\cite{Strominger:1990pd,Candelas:1990pi,Ishiguro:2020nuf,Ishiguro:2021ccl,Baur:2020yjl,Nilles:2021glx}.)
Also, they were used in flavor model building \cite{Ding:2021,Ding:2021iqp}.
Here, we  construct the orbifold twists as elements of $Sp(2g,\mathbb{Z})$, 
and modular transformation behavior of wave functions.
Then, we study zero-modes in $T^6/\mathbb{Z}_N$ orbifolds as well as  $T^4/\mathbb{Z}_N$ orbifolds.

The rest of our paper is organized as follows.
In section 2, we review massless spinor modes on $T^6$, and bosonic spectra.
In section 3, we give a brief review of 6D lattices leading to $T^6/\mathbb{Z}_N$ with 4D $\mathcal{N}=1$ SUSY.
We study magnetized  $T^6/\mathbb{Z}_7$, and $T^6/\mathbb{Z}_{12}$ models in 
section 4 and 5.
In section 6, we discuss the tachyon-free condition.
Section 7 is our conclusion. In Appendix A, we present results on magnetized $T^4/\mathbb{Z}_N$ orbifolds with $SO(8)$ Lie lattice. In Appendix B, we derive the transformations of zero-mode wave functions under $Sp(6,\mathbb{Z})$.

\section{Magnetized $T^6$ model}
First we consider magnetized D-brane models with $T^6$ compactification. We review the Dirac operator and the Dirac equation to introduce fermion zero-modes on magnetized $T^6$ \cite{Cremades:2004wa, Antoniadis:2009bg}.

\subsection{Dirac equation on magnetized $T^6$}
In order to find wave functions on magnetized $T^6$, we construct the Dirac operator on the six dimensional torus $T^6 \simeq \mathbb{C}^3 /\Lambda$, where $\Lambda$ is a lattice spanned by six basis vectors ${e}^{\prime}_{i}$, ($i = 1, 2, 3, 4, 5, 6$) defined by
\begin{align}
{e}^{\prime}_{1} &= 2\pi R \vec{e}_1 = 2\pi R 
\begin{bmatrix}
1 \\
0 \\
0 \\
\end{bmatrix},
{e}^{\prime}_{2} = 2\pi R \vec{e}_2 = 2\pi R 
\begin{bmatrix}
0 \\
1 \\
0 \\
\end{bmatrix},
{e}^{\prime}_{3} = 2\pi R \vec{e}_3 = 2\pi R 
\begin{bmatrix}
0 \\
0 \\
1 \\
\end{bmatrix} ,
\notag \\
{e}^{\prime}_{4} &= 2\pi R \vec{e}_4 = 2\pi R 
\begin{bmatrix}
\omega_1 \\
\omega_4 \\
\omega_6 \\
\end{bmatrix},
{e}^{\prime}_{5} = 2\pi R \vec{e}_5 = 2\pi R 
\begin{bmatrix}
\omega_4 \\
\omega_2 \\
\omega_5 \\
\end{bmatrix},
{e}^{\prime}_{6} = 2\pi R \vec{e}_6 = 2\pi R 
\begin{bmatrix}
\omega_6 \\
\omega_5 \\
\omega_3 \\
\end{bmatrix}.
\end{align}
Here, $R (>0)$ denotes the scale factor and $\omega_i \in \mathbb{C}$ characterize the shape of $\Lambda$.
By factoring out $R$, we defined vectors $\vec{e}_i$.
Here we focus on the six basis vectors corresponding to simply laced root lattices. 

Also, we define real coordinates $x^i , y^i$, ($i = 1, 2, 3$) along the lattice vectors on $T^6$. They are related to complex coordinates $\vec{Z} = (Z^1 , Z^2 , Z^3)$ of $\mathbb{C}^3$ by
\begin{align}
\label{eq: complex_coordinates}
\vec{Z}
=
\begin{bmatrix}
Z^1 \\
Z^2 \\
Z^3 \\
\end{bmatrix}
=
2\pi R 
\left(
\begin{bmatrix}
x^1 \\
x^2 \\
x^3 \\
\end{bmatrix}
+ 
\begin{bmatrix}
\omega_1 & \omega_4 & \omega_6 \\
\omega_4 & \omega_2 & \omega_5 \\
\omega_6 & \omega_5 & \omega_3 \\
\end{bmatrix}
\begin{bmatrix}
y^1 \\
y^2 \\
y^3 \\
\end{bmatrix}
\right)
=2\pi R (\vec{x} + \Omega \vec{y})
=2\pi R \vec{z},
\end{align} 
where we identify $\vec{z} = \vec{x} + \Omega \vec{y}$ as complex coordinates on $T^6$ and
\begin{align}
\Omega
=
\begin{bmatrix}
\omega_1 & \omega_4 & \omega_6 \\
\omega_4 & \omega_2 & \omega_5 \\
\omega_6 & \omega_5 & \omega_3 \\
\end{bmatrix},
\end{align}
is complex structure moduli. 
We are interested in symmetric moduli $\Omega^T = \Omega$, thus the actions of $Sp(6, \mathbb{Z})$ modular group can be consistently seen. 
Then we will discuss how to realize theories on $T^6/\mathbb{Z}_7$ and $T^6/\mathbb{Z}_{12}$.

Here, $\Omega$ is not necessarily an element of the Siegel upper-half plane $\mathcal{H}^3$ defined as\cite{Siegel},
\begin{align}
\label{eq: Siegel_uhs_3}
\mathcal{H}_3 = \{ \Omega \in GL(3, \mathbb{C}) | \Omega^T = \Omega , Im \Omega > 0  \}.
\end{align}
We will see that zero-modes of all positive chirality $(+,+,+)$ are well-defined if $N\Omega \in \mathcal{H}^3$, where $N$ is a $3\times 3$ integer matrix called flux and we will define latter. 

Then, we define the Dirac operator to write down the Dirac equation on magnetized $T^6$.
The K\"ahler metric on $\mathbb{C}^3$ is defined as,
\begin{align}
\label{eq: C3_Kahler}
ds^2 = 2H_{i\bar{j}} dZ^{i} d\bar{Z}^{\bar{j}},
\end{align}
where $H_{i\bar{j}} = \frac{1}{2} \delta_{i, \bar{j}}$ and $i, j =1, 2, 3$.

The Gamma matrices on $\mathbb{C}^3$ are defined as
\begin{align}
\begin{aligned}
\Gamma^{Z^{1}} &= \sigma^{Z} \otimes \sigma^3 \otimes \sigma^3, \quad
\Gamma^{\bar{Z}^{\bar{1} } } =  \sigma^{\bar{Z}} \otimes \sigma^3 \otimes \sigma^3, \\
\Gamma^{Z^{2}} &= {\bf{1}}_2 \otimes \sigma^{Z} \otimes \sigma^3,  \quad
\Gamma^{\bar{Z}^{\bar{2} } } = {\bf{1}}_2 \otimes \sigma^{\bar{Z}} \otimes \sigma^3, \\
\Gamma^{Z^{3}} &= {\bf{1}}_2 \otimes {\bf{1}}_2 \otimes \sigma^{Z},  \quad
\Gamma^{\bar{Z}^{\bar{3} } } = {\bf{1}}_2 \otimes  {\bf{1}}_2 \otimes \sigma^{\bar{Z}},
\end{aligned}
\end{align}
where ${\bf{1}}_2$ is the $2 \times 2$ unit matrix and $\sigma^i$ are Pauli matrices,
\begin{align}
{\bf{1}}_2 =
\begin{bmatrix}
1 & 0 \\
0 & 1 \\
\end{bmatrix},
\sigma^1 = 
\begin{bmatrix}
0 & 1 \\
1 & 0 \\
\end{bmatrix},
\sigma^2 = 
\begin{bmatrix}
0 & -i \\
i & 0 \\
\end{bmatrix},
\sigma^3 = 
\begin{bmatrix}
1 & 0 \\
0 & -1 \\
\end{bmatrix},
\end{align}
\begin{align}
\sigma^Z 
= 
\sigma^1 + i \sigma^2
=
\begin{bmatrix}
0 & 2 \\
0 & 0 \\
\end{bmatrix},
\sigma^{\bar{Z}} 
= 
\sigma^1 - i \sigma^2
=
\begin{bmatrix}
0 & 0 \\
2 & 0 \\
\end{bmatrix}.
\end{align}
Then one can find the K\"ahler metric on $T^6$ 
\begin{align}
\label{eq: T6_metric}
ds^2 = 2 h_{i\bar{j}} dz^{i} d\bar{z}^{\bar{j}}, \qquad 
h_{i\bar{j}} = (2\pi R)^2 H_{i\bar{j}}.
\end{align}
On the other hand, the Gamma matrices $\Gamma^{z^{i}}, \Gamma^{\bar{z}^{\bar{i} } } $ on complex coordinates of $T^6$ are as follows,
\begin{align}
\Gamma^{z^{i} } &= \frac{1}{2\pi R} \Gamma^{Z^{i} },\quad   
\Gamma^{\bar{z}^{\bar{i} } } = \frac{1}{2\pi R} \Gamma^{\bar{Z}^{\bar{i} } }, 
\end{align}
where $i = 1, 2, 3$. Then we obtain the following anti-commutative relations called the Dirac algebra (or Clifford algebra),
\begin{align}
\begin{aligned}
\{ \Gamma^{z^{i} } , \Gamma^{z^{j} }  \} &= \{ \Gamma^{\bar{z}^{\bar{i} } } , \Gamma^{\bar{z}^{\bar{j} } } \} = 0, \\
\{ \Gamma^{z^{i} } , \Gamma^{\bar{z}^{\bar{j} } }  \} &= 2h^{i\bar{j}}.
\end{aligned}
\end{align}
We define the chirality operator $\Gamma^5$ by 
\begin{align}
\Gamma^5 = \sigma^3 \otimes \sigma^3 \otimes \sigma^3
=
{\rm diag}[+,-,-,+,-,+,+,-].
\end{align}
We can write the Dirac operator on $T^6$ by the Gamma matrices
\begin{align}
i\slashed{D} 
&\equiv i (\Gamma^{z^{j}} D_{z^{j}} + \Gamma^{\bar{z}^{\bar{j} } } \bar{D}_{\bar{z}^{\bar{j}}} ) \notag \\
&= \frac{i}{\pi R}
\begin{bmatrix}
D_{2,3} & D_1 \\
\bar{D}_{\bar{1}} & D_{2,3}
\end{bmatrix},
\end{align}
where $D_{z^{j}}, \bar{D}_{\bar{z}^{\bar{j}}}$ are covariant derivatives and fermion are coupled to $U(1)$ gauge field with unit charge ($q=1$),
\begin{align}
\begin{aligned}
D_{z^{j}} = \partial_{z^j} - i A_{z^j}, \\
\bar{D}_{\bar{z}^{\bar{j}}} = \partial_{\bar{z}^{\bar{j}}} - i A_{\bar{z}^{\bar{j}}}.
\end{aligned}
\end{align}
Operators $D_{2,3}$ and $D_1$ are written by $D_{z^i}, \bar{D}_{\bar{z}^{\bar{i}}}$,
\begin{align}
D_{2,3} 
=
\begin{bmatrix}
0 & D_{z^3} & D_{z^2} & 0 \\
\bar{D}_{\bar{z}^{\bar{3}}} & 0 & 0 & -D_{z^2} \\
\bar{D}_{\bar{z}^{\bar{2}}} & 0 & 0 & D_{z^3} \\
0 & -\bar{D}_{\bar{z}^{\bar{2}}} & \bar{D}_{\bar{z}^{\bar{3}}} & 0 \\
\end{bmatrix},
D_1
=
\begin{bmatrix}
D_{z^1} & 0 & 0 & 0 \\
0 & -D_{z^1} & 0 & 0 \\
0 & 0 & -D_{z^1} & 0 \\
0 & 0 & 0 & D_{z^1} \\
\end{bmatrix}.
\end{align}

\subsection{Background magnetic flux and F-term condition}
In this subsection, we introduce background magnetic flux $F$ on $T^6$ \cite{Antoniadis:2009bg},
\begin{align}
F = \frac{1}{2} (p_{xx})_{ij} dx^{i} \wedge dx^{j} + \frac{1}{2} (p_{yy})_{ij} dy^{i} \wedge dy^{j} + (p_{xy})_{ij} dx^{i} \wedge dy^{j}.
\end{align}
In terms of complex coordinates $z^i$ we get,
\begin{align}
F = \frac{1}{2} (F_{zz})_{ij} dz^{i} \wedge dz^{j} + \frac{1}{2} (F_{\bar{z}\bar{z}})_{ij} d\bar{z}^{i} \wedge d\bar{z}^{j} + (F_{z\bar{z}})_{ij} (idz^{i} \wedge d\bar{z}^{j}),
\end{align}
where 
\begin{align}
\begin{aligned}
(F_{zz})_{ij} &= (\bar{\Omega} - \Omega)^{-1} (\bar{\Omega} p_{xx} \bar{\Omega} + p_{yy} +p^{T}_{xy} \bar{\Omega} -\bar{\Omega} p_{xy})(\bar{\Omega} - \Omega)^{-1}, \\
(F_{\bar{z}\bar{z}})_{ij} &= (\bar{\Omega} - \Omega)^{-1} ({\Omega} p_{xx} {\Omega} + p_{yy} +p^{T}_{xy} {\Omega} -{\Omega} p_{xy})(\bar{\Omega} - \Omega)^{-1}, \\
 (F_{z\bar{z}})_{ij} &= i (\bar{\Omega} - \Omega)^{-1} (\bar{\Omega} p_{xx} {\Omega} + p_{yy} +p^{T}_{xy} {\Omega} -\bar{\Omega} p_{xy})(\bar{\Omega} - \Omega)^{-1}.
\end{aligned}
\end{align}

We consider 10D $\mathcal{N} = 1$ supersymmetric Yang-Mills theory. Hermitian Yang-Mills equation \cite{Cremades:2004wa} to conserve SUSY  imposes a F-flat condition. That is the magnetic flux has to be $(1, 1)$-form, thus $F_{zz}=
F_{\bar{z}\bar{z}} = 0$. We will call it F-term condition, and it is equivalent to
\begin{align}
{\Omega} p_{xx} {\Omega} + p_{yy} +p^{T}_{xy} {\Omega} -{\Omega} p_{xy} = 0.
\end{align}

We can rewrite the magnetic flux as follows,
\begin{align}
\begin{aligned}
F 
&= (F_{z\bar{z}})_{ij} (idz^{i} \wedge d\bar{z}^{j}) 
&= i (p_{xx} \Omega -p_{xy}) (\bar{\Omega} - \Omega)^{-1} (idz^{i} \wedge d\bar{z}^{j}).
\end{aligned}
\end{align}
For simplicity, we assume $p_{xx} = p_{yy} = 0$.
Then we find 
\begin{align}
(F_{z\bar{z}})_{ij} = -i (p_{xy} (\bar{\Omega} - \Omega)^{-1})_{ij}.
\end{align}
From the F-term condition and the symmetry ($\Omega^T = \Omega$), one obtains
\begin{align}
p^{T}_{xy} \Omega = \Omega p_{xy} = (p^{T}_{xy} \Omega)^{T}.
\end{align}
From the Dirac quantization condition, the flux is written by an integer matrix $N$ as,
\begin{align}
p_{xy} = 2\pi N^T.
\end{align}
We just call $N$ as flux.
In summary, background magnetic flux $F$ is given by 
\begin{align}
\label{eq: background_F}
F = \pi (N^T (Im \Omega)^{-1})_{ij} (idz^i \wedge d\bar{z}^j),
\end{align}
and F-term condition is given by $(N\Omega)^T = N\Omega$.

\subsubsection{Gauge potential}
We find the gauge potential which corresponds to $F$ in eq.(\ref{eq: background_F}) as
\begin{align}
A (\vec{z}, \vec{\bar{z}}) 
&= \pi Im\left(N (\vec{\bar{z}} + \vec{\bar{\zeta}}) (Im \Omega)^{-1}) d\vec{z} \right)\notag \\
&= -\frac{i\pi}{2} \left( N (\vec{\bar{z}} + \vec{\bar{\zeta}}) (Im \Omega)^{-1}\right)_i dz^i + \frac{i\pi}{2} \left( N (\vec{z} + \vec{\zeta}) (Im \Omega)^{-1} \right)_i d\bar{z}^i \notag \\
&\equiv A_{z^i} dz^i + A_{\bar{z}^{\bar{i}}} d\bar{z}^i,
\end{align}
where $\vec{\zeta}$ is the Wilson line. 
The boundary conditions of gauge potential on $T^6$ are
\begin{align}
\label{eq: bcs_gauge_potential}
A(\vec{z} + \vec{e}_k) &= A(\vec{z}) + d\xi_{\vec{e}_k} (\vec{z}), \notag \\
A(\vec{z} + \Omega \vec{e}_k) &= A(\vec{z}) + d\xi_{\Omega \vec{e}_k} (\vec{z}),
\end{align}
where $\vec{e}_k$,$(k = 1, 2, 3)$ are 3 dimensional standard Euclidean unit vectors and
\begin{align}
\xi_{\vec{e}_k} (\vec{z}) &= \pi (N^T (Im \Omega)^{-1} Im(\vec{z}+\vec{\zeta}))_k, \notag \\
\xi_{\Omega\vec{e}_k} (\vec{z}) &= \pi Im (N \bar{\Omega} (Im \Omega)^{-1} (\vec{z}+\vec{\zeta}))_k.
\end{align}
In this paper, we assume that the Wilson line is vanishing, $\vec{\zeta}=\vec{0}$.

\subsection{The Dirac equation}
We introduce fermion massless modes (zero-modes) on magnetized $T^6$ which satisfy the following Dirac equation:
\begin{align}
i \slashed{D} \Psi (\vec{z} , {\vec{\bar{z}}}) =0,
\end{align}
where $\Psi (\vec{z} , {\vec{\bar{z}}})$ is an $8$ components spinor,
\begin{align}
\Psi (\vec{z} , {\vec{\bar{z}}}) = [\psi^{\prime}_{1+}, \psi_{3-}, \psi_{2-}, \psi_{1+}, \psi_{1-}, \psi_{2+}, \psi_{3+}, \psi^{\prime}_{1-}]^T.
\end{align}
$\psi_{j+}$ and $\psi_{j-}$ denote the positive and negative chirality components respectively. In particular, $\psi^{\prime}_{1+}$ denote that all chirality on $2D$ spinors are positive. That is, when we define Majorana-Weyl spinor on each complex planes as $(+,-)$, $\Psi (\vec{z} , {\vec{\bar{z}}})$ are given by
\begin{align}
\Psi (\vec{z} , {\vec{\bar{z}}}) 
\equiv
\begin{bmatrix}
+ \\
- \\
\end{bmatrix} 
\otimes
\begin{bmatrix}
+ \\
- \\
\end{bmatrix} 
\otimes
\begin{bmatrix}
+ \\
- \\
\end{bmatrix}, 
\end{align}
where $(+,+,+)$, $(+,-,-)$, $(-,+,-)$, $(-,-,+)$ correspond to $\psi^{\prime}_{1+}$, $\psi_{1+}$, $\psi_{2+}$, and $\psi_{3+}$.

From the definition of the Dirac operator $\slashed{D}$, we obtain the Dirac equation on each $\psi_i$ as follows:
\begin{align}
\begin{aligned}
\label{eq: full_Dirac}
D_{z^3} \psi_{3-} + D_{z^2} \psi_{2-} + D_{z^1} \psi_{1-} &= 0,  \\
\bar{D}_{\bar{z}^{\bar{3}}} \psi^{\prime}_{1+} - D_{z^2} \psi_{1+} - D_{z^1} \psi_{2+} &= 0,  \\
\bar{D}_{\bar{z}^{\bar{2}}} \psi^{\prime}_{1+} + D_{z^3} \psi_{1+} - D_{z^1} \psi_{3+} &= 0,\\
-\bar{D}_{\bar{z}^{\bar{2}}} \psi_{3-} + \bar{D}_{\bar{z}^{\bar{3}}} \psi_{2-} + D_{z^1} \psi^{\prime}_{1-} &= 0, \\
\bar{D}_{\bar{z}^{\bar{1}}} \psi^{\prime}_{1+} + D_{z^3} \psi_{2+} + D_{z^2} \psi_{3+} &= 0, \\  
-\bar{D}_{\bar{z}^{\bar{1}}} \psi_{3-} + \bar{D}_{\bar{z}^{\bar{3}}} \psi_{1-} - D_{z^2} \psi^{\prime}_{1-} &= 0, \\
-\bar{D}_{\bar{z}^{\bar{1}}} \psi_{2-} + \bar{D}_{\bar{z}^{\bar{2}}}\psi_{1-} + D_{z^3} \psi^{\prime}_{1-} &= 0,  \\
\bar{D}_{\bar{z}^{\bar{1}}}\psi_{1+} - \bar{D}_{\bar{z}^{\bar{2}}}\psi_{2+} + \bar{D}_{\bar{z}^{\bar{3}}} \psi_{3+} &= 0,
\end{aligned}
\end{align}
where 
\begin{align}
\begin{aligned}
D_{z^j} = \partial_{z^j} -\frac{\pi}{2} (N\vec{\bar{z}} (Im \Omega)^{-1})_{j}, \\
\bar{D}_{\bar{z}^{\bar{j}} } = \bar{\partial}_{\bar{z}^{\bar{j}} } +\frac{\pi}{2} (N\vec{z} (Im \Omega)^{-1})_{\bar{j}}.
\end{aligned}
\end{align}
Boundary conditions of $\Psi$ consistent with eq.(\ref{eq: bcs_gauge_potential}) are
\begin{align}
\begin{aligned}
\label{eq: boundary_conditions}
\Psi (\vec{z} + \vec{e}_k) &= e^{i\xi_{\vec{e}_k} (\vec{z})} \Psi (\vec{z}), \\
\Psi (\vec{z} + \Omega \vec{e}_k) &= e^{i\xi_{\Omega \vec{e}_k} (\vec{z})} \Psi (\vec{z}),
\end{aligned}
\end{align}
where $k=1,2,3$.

\subsection{Zero-modes on magnetized $T^6$}
We concentrate on zero-modes when the chirality $(+, +, +)$, satisfying $\bar{D}_{\bar{z}^{\bar{i}}} \psi_{1+}'=0$, hence eq.(\ref{eq: full_Dirac}) are solved with other spinor components are vanishing. We also require the boundary conditions eq.(\ref{eq: boundary_conditions}). The solution is given by\cite{Cremades:2004wa},
\begin{dmath}
\label{eq: zero-mode_1+}
\psi^{\vec{J}}_N (\vec{z}, \Omega)
=\mathcal{N} \cdot e^{i\pi (N\vec{z})^T (N Im\Omega)^{-1} \cdot Im(N \vec{z})} \cdot
\theta 
\begin{bmatrix}
\vec{J} N^{-1} \\
0 \\
\end{bmatrix}
(N\vec{z}, N\Omega)
= \mathcal{N} \cdot e^{i\pi (N \vec{z})^T (Im\Omega)^{-1} \cdot Im(\vec{z})} \cdot
\theta 
\begin{bmatrix}
\vec{J} N^{-1} \\
0 \\
\end{bmatrix}
(N\vec{z}, N\Omega),
\end{dmath}
where the Riemann-theta function with characteristics $\theta 
\begin{bmatrix}
\vec{J}N^{-1} \\
0 \\
\end{bmatrix}
(N\vec{z}, N\Omega)$ is defined by
\begin{align}
\theta
\begin{bmatrix}
\vec{a} \\
\vec{b} \\
\end{bmatrix}
(\vec{z} , \Omega')
= \sum_{\vec{m} \in \mathbb{Z}^3}
e^{\pi i (\vec{m} +\vec{a})^T \Omega' (\vec{m} +\vec{a})}e^{2\pi i (\vec{m} +\vec{a})^T (\vec{z} +\vec{b})},\quad \Omega' \in \mathcal{H}_3,\  \vec{a},\vec{b} \in \mathbb{R}^3.
\end{align}
The three components indices $\vec{J} \in \mathbb{Z}^3$ label the degeneracy of zero-modes. One can check the periodicity in the index $\vec{J}$,
\begin{align}
\psi^{\vec{J} +N^T \vec{e}_n}_{N} = \psi^{\vec{J}}_{N},
\end{align}
where $\vec{e}_n, (n=1,2,3)$ are 3 dimensional standard unit vectors. Thus, there are only $|\det N|$ of independent indices $\vec{J}$ and they lie inside the lattice $\Lambda_N$ spanned by  
\begin{equation}
\label{eq: Lambda_N}
    N^T \vec{e}_n,\quad (n=1,2,3).
\end{equation}
Here, note that zero-modes in eq.(\ref{eq: zero-mode_1+}) are well-defined if $\Omega' = N\Omega$ is an element of Siegel upper-half plane $\mathcal{H}_3$ defined in eq.(\ref{eq: Siegel_uhs_3}).
We stress here that $\Omega$ is not necessarily an element of $\mathcal{H}_3$.
We take the following normalization condition of wave functions,
\begin{align}
\int_{T^6} d^3 z d^3 \bar{z} \psi^{\vec{J}}_{N} (\psi^{\vec{K}}_{N})^{*} = (2^{3} \det (Im \Omega))^{-1/2} \delta_{\vec{J}, \vec{K}}.
\end{align}
Then the constant $\mathcal{N}$ is given by
\begin{align}
&\mathcal{N} = [Vol (T^6)]^{-1/2} (\det N)^{1/4}, \notag \\
&Vol (T^{6}) \propto \det (Im \Omega),
\end{align}
where $Vol (T^{6})$ represents volume of $T^6$ and is proportional to $\det (Im \Omega)$.

 Since ${\rm Im}\Omega^{\prime}$ is positive-definite, we have
\begin{align}
\det (N Im \Omega) = \det N \cdot \det (Im \Omega) >0. 
\end{align}
In the following, we will consider the case when $\det N > 0$ and $\det (Im \Omega) > 0$ are satisfied.

\subsubsection{Laplace operator}
Here, we confirm that the zero-mode wave functions in eq.(\ref{eq: zero-mode_1+}) are eigenfunctions of the Laplace operator on magnetized $T^6$.
Laplace operator is defined as follows
\begin{align}
\Delta = - {\frac{2}{(2\pi R)^2}} \sum_{j=1,2,3} \{ D_{z^j} , \bar{D}_{\bar{z}^{\bar{j}} } \}.
\end{align}

We focus on the spinor components which have positive chirality in the entire $6D$ compact space. When we act Laplace operator $\Delta$ on $\psi^{\prime}_{1+}$,
we find the eigenvalue equation,
\begin{align}
\label{eq: Laplace_eigen}
\Delta \psi^{\prime}_{1+} = 2(F_{z^1 {\bar{z}^{\bar{1}}}} + F_{z^2 {\bar{z}^{\bar{2}}}} + F_{z^3 {\bar{z}^{\bar{3}}}})\psi^{\prime}_{1+},
\end{align}
where we used the following commutation relations which are valid under the F-term condition,
\begin{align}
\begin{aligned}
[D_{z^{i}} , D_{z^{j}}] &= F_{z^{i} z^{j}} = 0, \\
[\bar{D}_{{\bar{z}}^{\bar{i}}} , \bar{D}_{{\bar{z}}^{\bar{j}}}] &= F_{{\bar{z}}^{\bar{i}} {\bar{z}}^{\bar{j}}} = 0, \\
[D_{z^{i}} , \bar{D}_{{\bar{z}}^{\bar{j}}}] &= F_{z^{i}{\bar{z}}^{\bar{j}}}.
\end{aligned}
\end{align}
Eq.(\ref{eq: Laplace_eigen}) shows that the eigenvalue is proportional to the trace of $F$. One can check that Laplacian $\Delta$ on a compact manifold is positive semi-definite. This means that $\psi_{1+}'$ is non-zero only if $F_{z^1 {\bar{z}^{\bar{1}}}} + F_{z^2 {\bar{z}^{\bar{2}}}} + F_{z^3 {\bar{z}^{\bar{3}}}} \geq 0$.

\subsection{Spectrum in the bosonic sector}
In this subsection, we show the mass spectrum of the 4D scalar fields which comes from dimensional reduction of 10D gauge boson. We will later use the obtained mass formula to discuss the stability and D-term SUSY condition of our magnetized orbifold models.

\subsubsection{Dimensional reduction of 10D SYM}
Here, we briefly review the dimensional reduction of 10D supersymmetric Yang-Mills theory (SYM). For simplicity, we consider $U(2)$ gauge group. Extension to $U(N)$ is straightforward. 
Our discussion is based on Ref. \cite{Conlon:2008qi} and appendix of Ref.\cite{Cremades:2004wa}. 
We assume compact space with no curvature such as $T^6$. 

The bosonic part of the action is 
\begin{align}
S_{YM} =  -\frac{1}{4g^2} \int d^{10} w\ 
{\rm Tr} \{ F^{MN} F_{MN}  \}, 
\end{align}
where $M$ and $N$ are the indices of ten-dimensional space-time, that is, $M, N \in \{ 0, 1, \cdots , 9\}$.
We take real orthogonal coordinate system $w^M$ with the following metric
\begin{align}
\eta_{MN} = {\rm diag} (-, +, +, \cdots , +, +).
\end{align} 
$F_{MN}$ is written as 
\begin{align}
F_{MN} &= \partial_M A_N -\partial_N A_M -i [ A_M , A_N ]. 
\end{align} 
Gauge boson $A_M$ is written as
\begin{align}
&A_M = B_M + W_M = B^{a}_M U_a + W^{ab}_M e_{ab},
\end{align}
where the elements of the Lie algebra of $U(2)$ are taken as $(U_a)_{ij} = \delta_{ai} \delta_{aj}$ and $(e_{ab})_{ij} = \delta_{ai} \delta_{bj}$ $(a \neq b)$. By noting $A^{\dagger}_M = A_M$, we see that $B^{a}_M$ is real and $(W^{ab}_M)^{*} = W^{ba}_{M}$. After the expansion, we obtain
\begin{align}
\label{eq: quadratic_LYM}
    \mathcal{L}_{YM} = - \frac{1}{2g^2} {\rm Tr} \left( D_M W_N D^M W^N - D_M W_N D^N W^M -iG_{MN} [W^M,W^N]  \right)+ \cdots,
\end{align}
where we have only shown quadratic terms of $W_M^{ab}$ explicitly, because we focus on mass terms. 3- and 4-point interactions are not relevant.
Here, we denote the field strength of Abelian direction by
\begin{align}
G_{MN} = \partial_M B_N -\partial_N B_M.
\end{align}
We also defined
\begin{equation}
D_M W_N = \partial_M W_N - i[B_M,W_N].
\end{equation}
Then we consider VEVs of Abelian constant magnetic fluxes,
\begin{align}
\label{eq: VEV}
&B^{a}_i (w) = \langle B^a_i \rangle (\eta) + C^a_i (w) 
, \notag \\
&W^{ab}_i (w) = 0 + \Phi_i^{ab}(w).
\end{align}
We consider $\langle B_i^{1} \rangle \neq \langle B_i^{2} \rangle$, then $U(2)$ gauge symmetry is broken to $U(1) \times U(1)$.
We take $w=(\xi,\eta)$ where $\xi$ denotes the real orthogonal coordinates of 4D space-time and $\eta$ denotes that of the compact space. Space-time indices are also decomposed as $M=(\mu, i)$ where $\mu =0,...,3$ and $i=4,..., 9$.

By substituting eq.(\ref{eq: VEV}) into eq.(\ref{eq: quadratic_LYM}), one obtains
\begin{dmath}
\label{eq: quadratic_LYM2}
\mathcal{L}_{YM} = \frac{i}{4g^2} (G^{a}_{ij} - G^{b}_{ij})(\Phi^{i, ab} \Phi^{j, ba} -\Phi^{j,ab} \Phi^{i, ba}) -\frac{1}{2g^2} [ (D_{\mu} \Phi^{ba}_i D^{\mu} \Phi^{i, ab}) + (\tilde{D}_i \Phi^{ba}_j \tilde{D}^i \Phi^{j, ab}) -2 (\tilde{D}_i W_{\mu}^{ba}) (D^{\mu} \Phi^{i ,ab}) - (\tilde{D}_i \Phi^{ba}_j) (\tilde{D}^j \Phi^{i, ab}) ] + \cdots,
\end{dmath}
where
\begin{equation}
    \tilde{D}_i W_j^{ab} = \partial_i W_j^{ab} -i(B_i^a - B_j^b)W_j^{ab}.
\end{equation}
If one takes the gauge fixing condition $\tilde{D}^i \Phi^{ab}_i = 0$, eq.(\ref{eq: quadratic_LYM2}) can be rewritten as
\begin{align}
\begin{aligned}
\label{eq: L_vector_mass}
    \mathcal{L}_{YM} =\frac{2i}{2g^2} \Phi^{j,ba} \langle G \rangle_{ij}^{ab} \Phi^{i,ab} + \frac{1}{2g^2} [ \Phi^{ba}_i (D_{\mu}  D^{\mu}\Phi^{i,ab}) 
    + \Phi_{j}^{ba} (\tilde{D}_i  \tilde{D}^i \Phi^{j,ab})] + \cdots,
\end{aligned}
\end{align}
where $\langle G \rangle_{ij}^{ab} = \langle G \rangle_{ij}^a - \langle G \rangle_{ij}^b$.

\subsubsection{4D scalar mass}
Now, we consider the Kaluza-Klein decomposition,
\begin{equation}
\label{eq: vector_KK}
    \Phi_i^{ab}(w) = \sum_n \varphi_{n,i}^{ab}(\xi) \otimes \phi_{n,i}^{ab}(\eta), 
\end{equation}
where the wave functions in the compact space satisfy 
\begin{equation}
\label{eq: Laplace_eigenfunction}
    \Delta \phi_{n,i}^{ab}(\eta) = \kappa_n^2 \phi_{n,i}^{ab}(\eta).
\end{equation}
Here, $\kappa_n^2 \geq 0$ denotes the eigenvalue of the Laplace operator $\Delta = - \tilde{D}_i \tilde{D}^i$. The index $n(=0,1,\cdots)$ denotes the excitation number and we set $\kappa_{n+1}^2 > \kappa_n^2$.
Substituting eq.(\ref{eq: vector_KK}) into eq.(\ref{eq: L_vector_mass}) and integration with respect to the internal coordinates $\eta$ yield
\begin{align}
    S_{4D} = S_{kinetic} + S_{mass} + \cdots,
\end{align}
where
\begin{equation}
\label{eq: real_kinetic}
    S_{kinetic}=   \sum_n \int d^4 \xi\  \delta_{ij} \varphi_n^{i,ba} (D_{\mu} D^{\mu}) \varphi_n^{j,ab},\quad   (a>b),
\end{equation}

\begin{equation}
\label{eq: real_mass}
    S_{mass} = \sum_n \int d^4 \xi \ \varphi_n^{j,ba} (2i \langle G \rangle_{ij}^{ab} - \kappa_n^2 \delta_{ij} ) \varphi_n^{i,ab},\quad  (a>b).
\end{equation}
Note that we have used the normalization condition,
\begin{equation}
    g^{-2} \int d^6 \eta\  {\phi_{m}^{i,ab}(\eta)}^* \phi_{n}^{i,cd}(\eta) =   \delta_{mn} \delta_{ac}\delta_{bd}.
\end{equation}

Next, we consider the coordinate transformation to the complex basis.
We define complex coordinates $\vec{z}$ as $z^j = (2\pi R)^{-1} (\eta^{2j+2} +i \eta^{2j+3})$ where $j=1,2,3$. Note that complex coordinates defined here are identified as those defined on the right-hand side of eq.(\ref{eq: complex_coordinates}). 
Then we have 
\begin{align}
\begin{aligned}
\varphi^{z^j,ab}(\vec{z}) &= \frac{1}{2\pi R}(\varphi^{2j+2,ab}(\eta) + i \varphi^{2j+3,ab}(\eta)), \\
\varphi^{\bar{z}^{\bar{j}},ab}(\vec{z}) &= \frac{1}{2\pi R}(\varphi^{2j+2,ab}(\eta) - i \varphi^{2j+3,ab}(\eta)).
\end{aligned}
\end{align}
Rewriting eq.(\ref{eq: real_kinetic}) in the new basis, we obtain 
\begin{align}
\begin{aligned}
    S_{kinetic} &=  \sum_n  \int d^4 \xi\  [h_{{i}{\bar{j}}} \varphi_n^{z^{i},ba} (D_{\mu} D^{\mu}) \varphi_n^{\bar{z}^{\bar{j}},ab} +
    h_{\bar{i} {j}} \varphi_n^{\bar{z}^{\bar{i}},ba} (D_{\mu} D^{\mu}) \varphi_n^{z^{j},ab}]
    \\ 
    &=\frac{(2\pi R)^2}{2} \sum_n  \int d^4 \xi\  [\varphi_n^{z^{j},ba} (D_{\mu} D^{\mu}) \varphi_n^{\bar{z}^{\bar{j}},ab} +
 \varphi_n^{\bar{z}^{\bar{j}},ba} (D_{\mu} D^{\mu}) \varphi_n^{z^{j},ab}],
 \end{aligned}
 \end{align}
 where we used metric of the form eq.(\ref{eq: T6_metric}).
 To get canonical kinetic term, we redefine the 4D scalar field as
 \begin{equation}
     \hat{\varphi}^{z^j,ab}=\frac{2\pi R}{\sqrt{2}} \varphi^{z^j,ab},\quad 
    \hat{\varphi}^{\bar{z}^{\bar{j}},ab} = \frac{2\pi R}{\sqrt{2}}     \varphi^{\bar{z}^{\bar{j}},ab}.
 \end{equation}
 Then we obtain 4D scalar mass term as
 \begin{align}
 \begin{aligned}
 \label{eq: 4D_scalar_mass_term}
    S_{mass} = \sum_n  \int d^4 \xi \ & \Big[ \hat{\varphi}_n^{z^{j},ba} \left( \frac{4i}{(2\pi R)^2} \langle G \rangle_{\bar{z}^{\bar{j}} z^j}^{ab} - \kappa_n^2 \right) \hat{\varphi}_n^{\bar{z}^{\bar{j}},ab} 
    \\
     +& \hat{\varphi}_n^{\bar{z}^{\bar{j}},ba}  \left(\frac{4i}{(2\pi R)^2} \langle G \rangle_{z^j \bar{z}^{\bar{j}}}^{ab} - \kappa_n^2  \right) \hat{\varphi}_n^{z^{j},ab} \Big].
\end{aligned}
\end{align}
If we assume $\langle G^{b=2} \rangle = 0$, scalar modes have only $U(1)_{a=1}$ charge and  we obtain 4D scalar mass formula in $U(1)$ gauge theory which we have been focusing. Eq.(\ref{eq: 4D_scalar_mass_term}) is realizes the results presented in Ref.\cite{Cremades:2004wa} in the case of $T^2$ compactification.

\section{$T^6/\mathbb{Z}_N$ orbifolds}

\subsection{6D lattices}

Here, we give a brief review on the $T^6/\mathbb{Z}_N$ orbifolds.
The $\mathbb{Z}_N$ twists preserving 4D $\mathcal{N}=1$ SUSY were studied in Refs.~\cite{Dixon:1985jw,Dixon:1986jc}, and 
also are shown in the second column of Table \ref{tab:lattice}, where 
eigenvalues of the orbifold twist are written by $e^{2\pi i k_i/N}$ $(i=1,2,3)$.
The SUSY condition requires 
\begin{align}
k_1 + k_2 +k_3 = 0~~({\rm mod}~N).
\end{align}
We divide the 6D flat space by a 6D lattice $\Lambda_6$ to construct  
the torus $T^6$.
We divide $T^6$ by the $\mathbb{Z}_N$ twist so as to obtain the $T^6/\mathbb{Z}_N$ orbifold.
Hence, the 6D lattice must have the $\mathbb{Z}_N$ symmetry.
We use Lie lattices with dimensions $D \leq 6$ and combine them to construct the 6D lattice $\Lambda_6$.
The $\mathbb{Z}_N$ twist corresponds to the Coxeter element $C$ of Lie root lattice 
\cite{Markushevich:1986za,Ibanez:1987pj,Katsuki:1989bf,Kobayashi:1991rp,Lust:2005dy,Lust:2006zg}.
For example, the Coxeter element $C$ of $SU(N)$  root lattice has the $\mathbb{Z}_N$ symmetry, i.e. $C^N=1$.
In particular, we use Lie root lattices with even dimensions, where we define complex coordinates and introduce magnetic 
fluxes.
We also use the $SU(2)^2$  lattice, but they are not always orthogonal to each other.
To represent $T^2/\mathbb{Z}_2$, we denote its lattice by $SU(2)^2$, because the product of their Coxeter elements
is the $\mathbb{Z}_2$ twist in two dimensions.
These root lattices are shown in Table \ref{tab:lattice}.
In the table, $SU(3)^{[2]}$ and $SO(8)^{[2]}$ denote use of 
generalized Coxeter elements including $\mathbb{Z}_2$ outer automorphisms of $SU(3)$ and $SO(8)$ Lie algebras, respectively.
In addition, $SO(8)^{[3]}$ denotes use of generalized Coxeter element of $SO(8)$ including $\mathbb{Z}_3$ outer automorphism.

\begin{table}[h]
  \centering
  \begin{tabular}{|c|c|c|}
    \hline
Orbifold & twists $(k_1,k_2,k_3)/N$ & lattice \\
\hline
$T^6/\mathbb{Z}_3$ & (1,1,-2)/3 & $SU(3)^3$ \\ 
$T^6/\mathbb{Z}_4$ & (1,1,-2)/4 & $(SO(4)^{[2]})^2\times SU(2)^2$\\
$T^6/\mathbb{Z}_{6-I}$ & (1,1,-2)/6 &$(SU(3)^{[2]})^2\times SU(3)$ \\
$T_6/\mathbb{Z}_{6-II}$ & (1,2,-3)/6 &$SU(3)^{[2]}\times SU(3) \times SU(2)^2$, ~~$SO(8)\times SU(3)$ \\
$T^6/\mathbb{Z}_7$ & (1,2,-3)/7 & $SU(7)$ \\
$T^6/\mathbb{Z}_{8-I}$ & (1,2,-3)/8 & $(SO(8)^{[2]})\times SO(4)^{[2]}$ \\
$T^6/\mathbb{Z}_{8-II}$ &(1,3,-4)/8 & $(SO(8)^{[2]})\times SU(2)^2$ \\
$T^6/\mathbb{Z}_{12-I}$ &(1,4,-5)/12 & $E_6$ \\
$T^6/\mathbb{Z}_{12-II}$ & (1,5-6)/12 & $(SO(8)^{[3]})\times SU(2)^2$   \\
\hline
\end{tabular}
\caption{$T^6/\mathbb{Z}_N$ orbifolds and torus lattices.}
\label{tab:lattice}
\end{table}

The flavor structure originated from $T^2/\mathbb{Z}_N$ has been studied already 
in Refs.~\cite{Abe:2008sx,Abe:2015yva}.
Thus, here we study non-factorizable $T^4/\mathbb{Z}_N$ and $T^6/\mathbb{Z}_N$ orbifolds in 
Table \ref{tab:lattice}, 
in particular the numbers of zero-modes on these orbifolds with magnetic fluxes.

Non-factorizable $T^6/\mathbb{Z}_N$ orbifolds include 
the $T^6/\mathbb{Z}_7$ orbifold with  the $SU(7)$ root lattice and the $T^6/\mathbb{Z}_{12-I}$ orbifold 
with $E_6$ root lattice.
The zero-modes on such magnetized orbifolds are studied in next sections.
We reconstruct the orbifold twists as elements of $Sp(6,\mathbb{Z})$.
We study zero-modes by analyzing $Sp(6,\mathbb{Z})$ modular transformation behaviors of 
wave functions.
Similar analysis was carried out in magnetized $T^2/\mathbb{Z}_N$ orbifold models \cite{Kobayashi:2017dyu}.
Our analysis is extensions to $T^6/\mathbb{Z}_N$ orbifolds as well as $T^4/\mathbb{Z}_N$ orbifolds.
(See also Ref.~\cite{Kikuchi:2022psj}.)

Non-factorizable $T^4/\mathbb{Z}_N$ orbifolds include 
$T^4/\mathbb{Z}_6$ in $T^6/\mathbb{Z}_{6-II}$,  
$T^4/\mathbb{Z}_8$ in $T^6/\mathbb{Z}_{8-II}$ and  $T^6/\mathbb{Z}_{8-II}$,  
and 
$T^4/\mathbb{Z}_{12}$ in $T^6/\mathbb{Z}_{12-II}$.
All of them use the $SO(8)$ root lattice.
However, we cannot introduce magnetic fluxes on 
$T^4/\mathbb{Z}_6$ and  $T^4/\mathbb{Z}_{12}$.
Its reason is explained in Appendix A.
Also Appendix A shows zero-modes of magnetized $T^4/\mathbb{Z}_8$ orbifold models.

\subsection{$\mathbb{Z}_N$ twists for $T^6/\mathbb{Z}_N$}

6D lattices have the modular symmetry, that is, the basis transformation of the basis vectors.
We find some of the aforementioned Coxeter and generalized Coxeter elements can be expressed as $Sp(6,\mathbb{Z})$ modular transformations.

The symplectic modular group $Sp(6, \mathbb{Z})$ is given by the set of $6\times 6$ integer matrices,
\begin{align}
\gamma
=
\begin{bmatrix}
A & B \\
C & D \\
\end{bmatrix},
\end{align}
satisfying
\begin{align}
\gamma J \gamma^T = J,\quad 
J
=
\begin{bmatrix}
0 & {\bf{1}}_3 \\
-{\bf{1}}_3 & 0 \\
\end{bmatrix}.
\end{align}
$A, B, C$, and $D$ are $3\times 3$ integer matrices.
The modular transformation of the complex coordinates $\vec{z}$ and the complex structure moduli $\Omega$ under $\gamma$ are given by
\begin{align}
\Omega &\rightarrow (A\Omega +B)(C\Omega +D)^{-1}, \\
\vec{z} &\rightarrow {(C\Omega +D)^{-1}}^{T}\vec{z}.
\end{align}
Generators $S$, $T_i$, $(i=1,2,\cdots,5,6)$ are given by
\begin{align}
S 
=
\begin{bmatrix}
O & {\bf{1}}_{3} \\
-{\bf{1}}_{3} & O \\
\end{bmatrix} , 
T_{i} =
\begin{bmatrix}
{\bf{1}}_3 & B_{i} \\
O & {\bf{1}}_3 \\
\end{bmatrix},
\end{align}
where $B_i$ are symmetric matrices given by
\begin{align}
&B_1 = 
\begin{bmatrix}
1 & 0 & 0 \\
0 & 0 & 0 \\
0 & 0 & 0 \\
\end{bmatrix}
, \quad 
B_2 = 
\begin{bmatrix}
0 & 0 & 0 \\
0 & 1 & 0 \\
0 & 0 & 0 \\
\end{bmatrix}
,\quad 
B_3 = 
\begin{bmatrix}
0 & 0 & 0 \\
0 & 0 & 0 \\
0 & 0 & 1 \\
\end{bmatrix}
,
\\
&B_4 = 
\begin{bmatrix}
0 & 1 & 0 \\
1 & 0 & 0 \\
0 & 0 & 0 \\
\end{bmatrix}
, \quad 
B_5 = 
\begin{bmatrix}
0 & 0 & 0 \\
0 & 0 & 1\\
0 & 1 & 0 \\
\end{bmatrix}
,\quad 
B_6 = 
\begin{bmatrix}
0 & 0 & 1 \\
0 & 0 & 0 \\
1 & 0 & 0 \\
\end{bmatrix}.
\end{align}
We will consider modular $S$ and $T_i$ transformations of zero-modes on magnetized $T^6$.

As we will see in the next section, the $\mathbb{Z}_7$ twist on the $SU(7)$ lattice can be written by $ST_3T_4T_5$ satisfying $(ST_3 T_4 T_5)^7=1$.
The $\mathbb{Z}_{12}$ twist on the $E_6$ lattice can be written by $ST_1 T_2 T_3^{-1} T_5 T_6$ satisfying $(ST_1 T_2 T_3^{-1} T_5 T_6)^{12}=1$.

In $Sp(6, \mathbb{Z})$ modular group, we suppose symmetric moduli $\Omega$, and one can see that the lattice of $T^6/\mathbb{Z}_N$ can be found by $\Omega$.

In the following section, we represent lattice vectors $\vec{e}_i$ as following Euclidean basis representation
\begin{align}
\vec{e}_1
=
\begin{bmatrix}
1 \\
0 \\
0 \\
0 \\ 
0 \\
0 \\
\end{bmatrix}
,\quad
\vec{e}_2
=
\begin{bmatrix}
0 \\
1 \\
0 \\
0 \\ 
0 \\
0 \\
\end{bmatrix}
,\quad
\vec{e}_3
=
\begin{bmatrix}
0 \\
0 \\
1 \\
0 \\ 
0 \\
0 \\
\end{bmatrix}
, \quad
\vec{e}_4
=
\begin{bmatrix}
Re \omega_1 \\
Re \omega_4 \\
Re \omega_6 \\
Im \omega_1 \\
Im \omega_4 \\ 
Im\omega_6 \\
\end{bmatrix}
, \quad
\vec{e}_5
=
\begin{bmatrix}
Re \omega_4 \\
Re \omega_2 \\
Re \omega_5 \\
Im \omega_4 \\
Im \omega_2 \\ 
Im\omega_5 \\
\end{bmatrix}
,\quad
\vec{e}_6
=
\begin{bmatrix}
Re \omega_6 \\
Re \omega_5 \\
Re \omega_3 \\
Im \omega_6 \\
Im \omega_5 \\ 
Im\omega_3 \\
\end{bmatrix}
.
\end{align}
Then we can find what lattices of $T^6/\mathbb{Z}_N$ correspond to root lattices of Lie algebra.

\section{Non-factorizable orbifolds}
Here we perform counting of the zero-modes in magnetized $T^6/\mathbb{Z}_7$ and $T^6/\mathbb{Z}_{12}$ orbifold models.

\subsection{Magnetized $T^6/\mathbb{Z}_7$ orbifold}
First we study zero-modes on magnetized $T^6/\mathbb{Z}_7$ orbifold. To realize the orbifold, we focus on the following algebraic relation,
\begin{align}
(ST_3 T_4 T_5)^7 = {\bf{1}_6}.
\end{align}
This shows that $ST_3 T_4 T_5$ transformation can be identified as the $\mathbb{Z}_7$-twist. Thus, it is useful for constructing $T^6/\mathbb{Z}_7$ orbifold.
Under the transformation $ST_3 T_4 T_5$, the complex structure moduli $\Omega$ and complex coordinates $\vec{z}$ transform as,
\begin{align}
\begin{aligned}
&\Omega \rightarrow -(\Omega +B_3 +B_4 +B_5)^{-1}, \\
&\vec{z} \rightarrow -(\Omega +B_3 +B_4 +B_5)^{-1} \vec{z}.
\end{aligned}
\end{align}
Then one can verify that $ST_3 T_4 T_5$ invariant moduli $\Omega_7$ are given by 
\begin{align}
\Omega_7
=
\begin{bmatrix}
\omega_1 & \omega_4 & \omega_6 \\
\omega_4 & \omega_2 & \omega_5 \\
\omega_6 & \omega_5 & \omega_3 \\
\end{bmatrix} 
=
\begin{bmatrix}
-\frac{2}{\sqrt{7}} i & -\frac{1}{2} +\frac{\sqrt{7}}{14} i & \frac{i}{\sqrt{7}} \\
-\frac{1}{2} +\frac{\sqrt{7}}{14} i & -\frac{i}{\sqrt{7}} &  -\frac{1}{2} +\frac{3\sqrt{7}}{14} i \\
 \frac{i}{\sqrt{7}} &  -\frac{1}{2} +\frac{3\sqrt{7}}{14} i & -\frac{1}{2} -\frac{\sqrt{7}}{14} i \\
\end{bmatrix},
\end{align}
where we take the case when $\det (Im \Omega_7) > 0$.
This corresponds to the $SU(7)$ root lattice as shown in Figure 1.
\begin{figure}[htbp]
\begin{center}
\includegraphics[width=80mm]{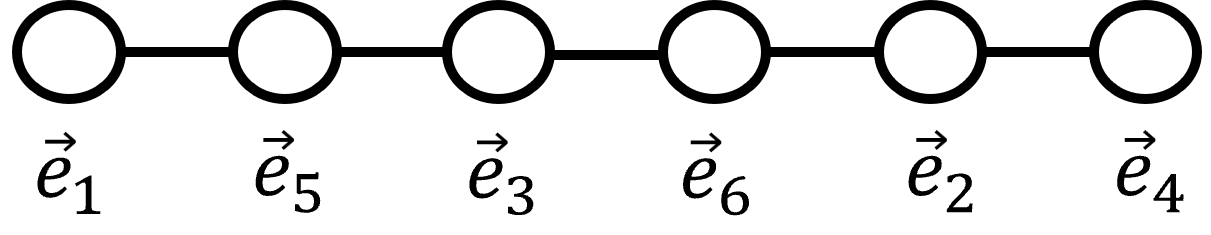}
\caption{The lattice of $T^6/\mathbb{Z}_7$}
\end{center}
\end{figure}

We note that the shape of flux $N$ is constrained as follows.
First, from the F-term condition $(N\Omega_7)^T = N\Omega_7$, $N$ is symmetric and parameterized as
\begin{align}
N = 
\begin{bmatrix}
n_{11} & n_{33} -n_{22} & n_{22} - n_{11} \\
n_{33} -n_{22} & n_{22} & n_{33} - n_{11} \\
n_{22} - n_{11} & n_{33} - n_{11} & n_{33} \\
\end{bmatrix},
\end{align}
where $n_{11}$, $n_{22}$, and $n_{33}$ are integers and we can see them as independent parameters.

Second, we consider the consistency with T-transformation. As we study in Appendix \ref{appendix: modular_T_zero}, the matrix $NB$ must be symmetric and its diagonal components are all even. From
\begin{align}
N(B_3 +B_4 +B_5) 
= 
\begin{bmatrix}
n_{33} -n_{22} & n_{22} & n_{33} - n_{11} \\
n_{22} & 2n_{33} - n_{11} - n_{22} & n_{33}+n_{22} - n_{11} \\
n_{33} - n_{11} & n_{33} +n_{22} - n_{11} & 2n_{33}-n_{11} \\
\end{bmatrix},
\end{align}
it is immediate that
\begin{align}
n_{11} \equiv n_{22} \equiv n_{33} \equiv 0 \quad \pmod2,
\end{align}
must be satisfied. As a result, we find that the background magnetic flux $F$ is invariant under the $\mathbb{Z}_7$ twist. 

We see that  $\det N$ is always a multiple of eight. We will analyze how many zero-modes with $\mathbb{Z}_7$ charges exist under the flux $N$.
\subsubsection{The number of zero-modes}
Here we analyze the number of zero-modes on magnetized $T^6/\mathbb{Z}_7$ with modular transformation. 
Noting that $\Omega_7$ satisfies $\Omega_7 = - (\Omega_7 +B_3+B_4+B_5)^{-1}$, zero-modes behave under the $ST_3 T_4 T_5$ transformation as,
\begin{align}
\begin{aligned}
 \psi^{\vec{J}}_{N}& (\Omega_7 \vec{z} , \Omega_7 )  \\
&= \frac{\sqrt{\det [-i(\Omega_7 +B_3 +B_4 +B_5)]}}{\sqrt{\det N}} \sum_{\vec{K} \in \Lambda_{N}} e^{2\pi i\vec{J}^{T} N^{-1} \vec{K}} e^{\pi i\vec{K}^{T} N^{-1} (B_3 +B_4 +B_5) \vec{K}} \psi^{\vec{K}}_{N} (\vec{z}, \Omega_7),
\end{aligned}
\end{align}
where $\Lambda_N$ is defined in eq.(\ref{eq: Lambda_N}).
We take a branch $\sqrt{t} > 0$ for $t > 0$. Then, it is found that 
\begin{dmath}
\sqrt{\det [-i(\Omega_7 +B_3 +B_4 +B_5)]}
=e^{-\pi i/4}.
\end{dmath}
We find the trace of $\rho(ST_3 T_4 T_5)$,
\begin{dmath}
tr \rho (ST_3 T_4 T_5)
= \frac{e^{-\pi i/4}}{\sqrt{\det N}} \sum_{\vec{K} \in \Lambda_{N}} e^{2\pi i\vec{K}^{T} N^{-1}\left({\bf{1}}_{3} +\frac{1}{2}(B_3 +B_4 +B_5)\right) \vec{K}}.
\end{dmath}
The determinant of $N$ is always a multiple of eight, $\det N = 8\ell, (\ell \in \mathbb{N})$. From numerical calculation in the region $|n_{ii}| \leq 400$, we obtained only three values of tr$\rho$,
\begin{align}
tr \rho = -1, +1, -\sqrt{7}i.
\end{align}
On the other hand, ${\rm tr}\rho$ can be expressed as
\begin{align}
tr \rho = \sum_{k = 0}^{6} n_k \gamma^k,\quad \gamma=e^{2\pi i/7},
\end{align}
where $n_k$ denotes the number of zero-modes which corresponds to the $\mathbb{Z}_7$ eigenvalue $\gamma^k$. 
Note that we also have $\sum_{k=0}^{6} n_k = \det N = 8\ell$.
In the simple case, the modes with $\gamma^0$ correspond to the $\mathbb{Z}_N$ invariant states.
However, we may embed the geometrical twist into the gauge sector.
Then, which state with $\gamma^k$ can survive through the $\mathbb{Z}_N$ projection depends on 
such gauge embedding.

First, we discuss the case when ${\rm tr}\rho = -\sqrt{7}i$. One can find that $[n_0, n_1, \cdots, n_5, n_6] = [1, 0, 0, 2, 0, 2, 2]$ reproduces ${\rm tr}\rho = -\sqrt{7}i$. Other possibilities are given by increasing each $n_k$ by $m \in \mathbb{Z}^+$ because $\sum_{k=0}^6 \gamma^k = 0$ holds. That is, $[n_0, n_1, \cdots, n_5, n_6] = [1+m, m, m, 2+m, m, 2+m, 2+m]$. 
As a result, $\det N$ is increased by $7m$ and we also observe that the minimal degeneracy number is given by $n_1, n_2$, and $n_4$ which are equal to $m$. Now, recall the fact that ${\rm det}N$ is a multiple of eight. We have the following equation
\begin{align}
8\ell = 7 + 7m = 7(m+1).
\end{align}
Since $7$ and $8$ are coprime, $m+1$ must be a multiple of eight. Thus we obtain possible values of $m$ as
\begin{align}
m = 7, 15, \cdots.
\end{align}
One can see that there are at least seven-generations when ${\rm tr}\rho = -\sqrt{7}i$.

Next we discuss the case tr $\rho = +1$. The first candidate is clearly $[n_0, n_1, \cdots, n_5, n_6] = [1, 0, 0, 0, 0, 0, 0]$ and the next one is $[n_0, n_1, \cdots, n_5, n_6] = [2, 1, 1, 1, 1, 1, 1]$.
By the similar discussion as for the case ${\rm tr}\rho = -\sqrt{7}i$, we find 
\begin{align}
\det N = 8\ell = 1 + 7m = 7(m-1) +8.
\end{align}
Then $m-1$ must be a multiple of eight,
\begin{align}
m = 1, 9, \cdots.
\end{align}
We can see that the minimal generation numbers are $1$, $9$ and so on. Therefore we cannot obtain three-generation model in the case of ${\rm tr}\rho = +1$.

Similarly, we cannot find three-generation models when ${\rm tr}\rho = -1$. 

In conclusion, there is no three-generation model on magnetized $T^6/\mathbb{Z}_7$. 

\subsection{Magnetized $T^6/\mathbb{Z}_{12}$ orbifold}

In this subsection, we focus on magnetized $T^6/\mathbb{Z}_{12}$ orbifold, whose twist is constructed by the following algebraic relation
\begin{align}
(ST_1 T_2 T^{-1}_3 T_5 T_6)^{12} = {\bf{1}}_6
\end{align}
In the following, we denote $ST_1 T_2 T^{-1}_3 T_5 T_6$ as $G$. We adopt the following complex structure moduli $\Omega = \Omega_{12}$ which are invariant under the $G$ transformation  
\begin{align}
\Omega_{12}
=
\begin{bmatrix}
-\frac{1}{2} -\frac{\sqrt{3}}{6} i & \frac{\sqrt{3}}{3} i & -\frac{1}{2} +\frac{\sqrt{3}}{6} i \\
\frac{\sqrt{3}}{3} i  & -\frac{1}{2} -\frac{\sqrt{3}}{6} i  & -\frac{1}{2}+\frac{\sqrt{3}}{6} i  \\
-\frac{1}{2} +\frac{\sqrt{3}}{6} i & -\frac{1}{2}+\frac{\sqrt{3}}{6} i  & \frac{1}{2}-\frac{\sqrt{3}}{6} i 
\end{bmatrix}
.
\end{align}
One can verify that this $T^6/\mathbb{Z}_{12}$ lattice corresponds to the $E_6$ root lattice as shown in Figure 2.  
\begin{figure}[htbp]
\begin{center}
\includegraphics[width=80mm]{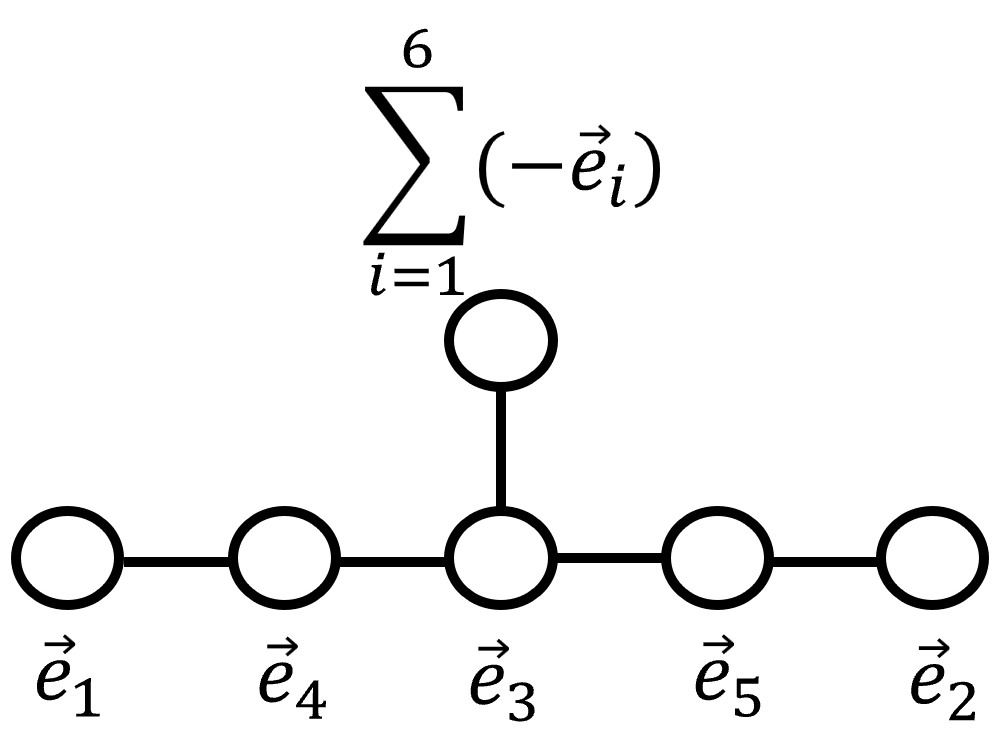}
\caption{The lattice of $T^6/\mathbb{Z}_{12}$}
\end{center}
\end{figure}

Shape of the flux $N$ is constrained. First, the F-term condition imposes a constraint, $(N\Omega_{12})^T = N\Omega_{12}$.
We also require the invariance of the flux under $G$ since the flux should be invariant for $\mathbb{Z}_{12}$-twist. Then the flux is symmetric. Also this includes the consistency with $T=T_1T_2T_3^{-1}T_5T_6$ which demands that $N(B_1 + B_2 -B_3 +B_5 +B_6)$ is symmetric and its diagonal elements are all even. One can find that fluxes of the form
\begin{align}
N
=
\begin{bmatrix}
n_{11} & n_{12} & n_{13} \\
n_{12}  & n_{11} & n_{13}  \\
n_{13} & n_{13} & n_{11} + n_{12} -2n_{13} \\ 
\end{bmatrix}
,
\end{align}
satisfy all requirements provided $n_{11} \equiv n_{12} \equiv n_{13} \pmod2$.

It is obvious that $G^2$ can be regarded as a $\mathbb{Z}_6$-twist. Also we have
\begin{align}
G^{12} = (G^{2})^6 = (G^{3})^4 = (G^{4})^3 = (G^{6})^2 = {\bf{1}}_6.
\end{align}
We will use this fact to count the numbers of  zero-modes. 

\subsubsection{Representation of modular transformation of zero-modes}
Modular transformation of wave function on magnetized $T^6/\mathbb{Z}_{12}$ is written as
\begin{dmath}
\psi^{\vec{J}}_{N} (\Omega_{12} \vec{z}, \Omega_{12}  )
=\frac{\sqrt{\det [-i (\Omega_{12} +B)]}}{\sqrt{\det N}} \sum_{\vec{K} \in \Lambda_N} e^{2\pi i\vec{J}^{T} N^{-1} \vec{K}} e^{\pi i \vec{K}^T N^{-1} B \vec{K}} \psi^{\vec{K}}_{N} (\vec{z}, \Omega_{12}),
\end{dmath}
where $B = B_1 +B_2 -B_3 +B_5 +B_6$. 
Here the representation itself is written by indices $\vec{K}_1$ and $\vec{K}_2$,
\begin{align}
\rho_{\vec{K}_1 \vec{K}_2 } (G) = \frac{e^{-\frac{\pi}{4} i}}{\sqrt{\det N}} e^{2\pi i \vec{K}^{T}_1 N^{-1} \vec{K}_2} e^{i\pi (\vec{K}^{T}_1 N^{-1} B \vec{K}_1)}.
\end{align}
Thus, trace of the representation $\rho (G)$ is given by
\begin{dmath}
{\rm tr} \rho(G) = \frac{e^{-\frac{\pi i}{4}}}{\sqrt{\det N}} \sum_{\vec{K} \in \Lambda_N} e^{2\pi i\vec{K}^{T} N^{-1} ({\bf{1}}_3 +B/2) \vec{K}}.
\end{dmath}
Then we immediately see following relation of tr $\rho(G^n)$ from property of modular transformation
\begin{align}
tr \rho (G^n) = \frac{e^{-\frac{\pi i}{4} n } }{(\det N)^{n/2}} \sum_{\vec{K}_1, \vec{K}_2, \cdots \vec{K}_n \in \Lambda_N} e^{2\pi i (\vec{K}^{T}_{1} N^{-1} \vec{K}_2 + \vec{K}^{T}_{2} N^{-1} \vec{K}_3 + \cdots + \vec{K}^{T}_{n} N^{-1} \vec{K}_1)}\notag \\
\cdot e^{\pi i (\vec{K}^{T}_{1} N^{-1} B \vec{K}_1 + \cdots + \vec{K}^{T}_{n} N^{-1} B \vec{K}_n)}.
\end{align}

\subsubsection{Zero-modes on magnetized $T^6/\mathbb{Z}_{12}$}
We show how to count zero-modes with $\mathbb{Z}_{12}$ charges. We denote the number of degenerated zero-modes corresponding to the $\mathbb{Z}_{12}$ eigenvalues $e^{\frac{k\pi}{6} i}$ $(k = 0,1, \cdots , 11)$ by $n_k$. 
Note that summation of all the degeneracy number is equal to $\det N$,
\begin{align}
\det N = \sum_{k=0}^{11} n_k.
\end{align}
Next we consider relations between tr$\rho (G^n)$ and coefficients $n_k$.
Since $G^{12} = {\bf{1}}_6$ and ${\rm tr}\rho(G)$ is summation of $\rho(G)$'s eigenvalues, we find
\begin{align}
{\rm tr} \rho (G) =  \sum_{k=0}^{11} n_k e^{\frac{k \pi}{6} i}.
\end{align}
We can represent ${\rm tr}\rho (G^n)$ as linear combinations of $n_k$ as,
 \begin{align}
{\rm tr}\rho(G^2) 
=
\sum_{k \equiv 0 (mod6)} n_k +e^{\frac{\pi}{3} i} \sum_{k \equiv 1 (mod6)} n_k +e^{\frac{2\pi}{3} i} \sum_{k \equiv 2 (mod6)} n_k \notag  \\ -\sum_{k \equiv 3 (mod6)} n_k +e^{\frac{4\pi}{3} i} \sum_{k \equiv 4 (mod6)} n_k  +e^{\frac{5\pi}{3} i} \sum_{k \equiv 5 (mod6)} n_k ,
\end{align}
\begin{align}
{\rm tr}\rho(G^3)  = \sum_{k \equiv 0 (mod4)} n_k +i  \sum_{k \equiv 1 (mod4)} n_k - \sum_{k \equiv 2 (mod4)} n_k -i \sum_{k \equiv 3 (mod4)} n_k,
\end{align}
\begin{align}
{\rm tr}\rho(G^4)  =  \sum_{k \equiv 0 (mod3)} n_k +e^{\frac{2\pi}{3}i} \sum_{k \equiv 1 (mod3)} n_k + e^{\frac{4\pi}{3}i} \sum_{k \equiv 2 (mod3)} n_k,
\end{align}
\begin{align}
{\rm tr}\rho(G^6)  =  \sum_{k \equiv 0 (mod2)} n_k -\sum_{k \equiv 1 (mod2)} n_k .
\end{align}
Then one can obtain the  number of zero-modes on magnetized $T^6/\mathbb{Z}_{12}$.
Here we conduct numerical calculation in the region $|n_{1i}| \leq 400$, and Tables 2 and 3 show the results with F-term condition.

\begin{table}[htb]
\centering
  \caption{The number of zero-modes on magnetized  $T^6/\mathbb{Z}_{12}$}
  \begin{tabular}{|l||c|r|c|r|c|r|c|r|c|r|c|r|c|r}  \hline
    ${\rm det}N$ & 4 & 8 & 12 & 16 & 20 & 24 & 24 & 28 & 32 & 32 & 36 & 36& 40 &$\cdots$ \\ \hline \hline
   ${\rm tr}\rho(G)$ & $-1$ & 1 & $-\sqrt{3}i$ & $-1$ & 1 & $-\sqrt{3}i$ & $-1$ & $-1$ & 1 & 1 & $-\sqrt{3}i$ & $-1$& $-1$  \\ \hline
   ${\rm tr}\rho(G^2)$ & 1 & $-1$ & $-\sqrt{3}i$ & 1 & $-1$ & $-\sqrt{3}i$ & 1 & 1 & $-1$ & $-1$ & $-\sqrt{3}i$ & 1 & 1 \\ \hline
   ${\rm tr}\rho(G^3)$ & 2 & 4 & 6 & 8 & 10 & 12 & 2 & 14 & 16 & 4 & 18 & 2 & 20\\ \hline
   ${\rm tr}\rho(G^4)$ & 1 & $-1$ & $\sqrt{3}i$ & 1 & $-1$ & $\sqrt{3}i$ & $-3$ & 1 & $-1$ & $-1$ & $\sqrt{3}i$ & $-3$ & 1\\ \hline
   ${\rm tr}\rho(G^6)$ & 4 & 8 & 12 & 16 & 20 & 24 & 4 & 28 & 32 & 8 & 36 & 4 & 40\\ \hline \hline
   $n_0$ & 1 & 2 & $\fbox{3}$ & 4 & 5 & 6 & 2 & 7 & 8 & 4 & 9 & $\fbox{3}$& 10\\ \hline
   $n_1$ & 0 & 0 & 0 & 0 & 0 & 0 & 2 & 0 & 0 & 2 & 0 & $\fbox{3}$ & 0\\ \hline
   $n_2$ & 0 & 1 & 0 & 1 & 2 & 1 & 2 & 2 & $\fbox{3}$ & $\fbox{3}$ & 2 & $\fbox{3}$ & $\fbox{3}$\\ \hline
   $n_3$ & 0 & 0 & 0 & 0 & 0 & 0 & 1 & 0 & 0 & 2 & 0 & 2& 0\\ \hline
   $n_4$ & 1 & 2 & $\fbox{3}$ & 4 & 5 & 6 & $\fbox{3}$ & 7 & 8 & 4 & 9 & 4 & 10\\ \hline
   $n_5$ & 0 & 0 & 0 & 0 & 0 & 0 & 2 & 0 & 0 & 2 & 0 & $\fbox{3}$  & 0\\ \hline
   $n_6$ & 1 & 0 & 1 & 2 & 1 & 2 & 2 & $\fbox{3}$ & 2 & 2 & $\fbox{3}$ & $\fbox{3}$ & 4\\ \hline
   $n_7$ & 0 & 0 & 0 & 0 & 0 & 0 & 2 & 0 & 0 & 2 & 0 & $\fbox{3}$ & 0 \\ \hline
   $n_8$ & 1 & 2 & $\fbox{3}$ & 4 & 5 & 6 & $\fbox{3}$ & 7 & 8 & 4 & 9 & 4 & 10\\ \hline
   $n_9$ & 0 & 0 & 0 & 0 & 0 & 0 & 1 & 0 & 0 & 2 & 0 & 2 & 0 \\ \hline
   $n_{10}$ & 0 & 1 & 2 & 1 & 2 & $\fbox{3}$ & 2 & 2 & $\fbox{3}$ & $\fbox{3}$ & 4 & $\fbox{3}$ & $\fbox{3}$ \\ \hline
   $n_{11}$ & 0 & 0 & 0 & 0 & 0 & 0 & 2 & 0 & 0 & 2 & 0 & $\fbox{3}$ & 0\\ \hline
  \end{tabular}
\end{table}
\begin{table}[htb]
\centering
  \caption{Continuation of Table 2}
  \begin{tabular}{|l||c|r|c|r|c|r|c|r|c|r|c|r|c|r}  \hline
    ${\rm det}N$  & 44 & 48 & 52 & 52 & 56 & 60 & 64 & 64 & 68 & 72 & 72 & 72 &$\cdots$ \\ \hline \hline
   ${\rm tr}\rho(G)$  & 1 & $-\sqrt{3}i$ & $-1$ & $-1$ & 1 & $-\sqrt{3}i$ & $-1$ & $-1$ & 1 & $-\sqrt{3}i$ & 1&$-\sqrt{3}i$\\ \hline
   ${\rm tr}\rho(G^2)$  & $-1$ & $-\sqrt{3}i$ & 1 & 1 & $-1$ & $-\sqrt{3}i$ & 1 & 1 & $-1$ & $-\sqrt{3}i$ & $-1$&$-\sqrt{3}i$ \\ \hline
   ${\rm tr}\rho(G^3)$  & 22 & 24 & 26 & 2 & 28 & 30 & 32 & 8 & 34&36&4&6\\ \hline
   ${\rm tr}\rho(G^4)$  & $-1$ & $\sqrt{3}i$ & 1 & 1 & $-1$ & $\sqrt{3}i$ & 1 & 1 & $-1$&$\sqrt{3}i$&3&$-3\sqrt{3}i$\\ \hline
   ${\rm tr}\rho(G^6)$  & 44 & 48 & 52 & 4 & 56 & 60 & 64 & 16 & 68 & 72&8&12\\ \hline \hline
   $n_0$  & 11 & 12 & 13 & 5 & 14 & 15 & 16 & 8 & 17&18&8&8 \\ \hline
   $n_1$   & 0   & 0 & 0 & 4 & 0 & 0 & 0 & 4 & 0&0&5&4 \\ \hline
   $n_2$  & 4 & $\fbox{3}$ & 4 & 4 & 5 & 4 & 5 & 5 & 6&5&6&6 \\ \hline
   $n_3$  & 0 & 0 & 0 & 4 & 0 & 0 & 0 & 4 & 0&0&6&5 \\ \hline
   $n_4$ & 11 & 12 & 13 & 5 & 14 & 15 & 16 & 8 & 17&18&7&7 \\ \hline
   $n_5$ & 0 & 0 & 0 & 4 & 0 & 0 & 0 & 4 & 0&0&5&6 \\ \hline
   $n_6$  & $\fbox{3}$ & 4 & 5 & 5 & 4 & 5 & 6 & 6 & 5 &6&6&6\\ \hline
   $n_7$ & 0 & 0 & 0 & 4 & 0 & 0 & 0 & 4 & 0&0 & 5&4\\ \hline
   $n_8$  & 11 & 12 & 13 & 5 & 14 & 15 & 16 & 8 & 17&18&7&9 \\ \hline
   $n_9$ & 0 & 0 & 0 & 4 & 0 & 0 & 0 & 4 & 0 & 0&6&5\\ \hline
   $n_{10}$  & 4 & 5 & 4 & 4 & 5 & 6 & 5 & 5 & 6 &7&6&6\\ \hline
   $n_{11}$ & 0 & 0 & 0 & 4 & 0 & 0 & 0 & 4 & 0 & 0&5&6\\ \hline
  \end{tabular}
\end{table}
\newpage

\section{D-term condition}
In this section, we study D-term condition by computing 4D scalar mass spectrum.
 Three-generation models satisfying the F-term condition were discussed in Sec. 4. 
  However, if tachyonic modes appeared in the models, we would treat unstable vacuum. 
If a model satisfies the D-term condition, the model is stable and phenomenologically attractive.

We have introduced $3 \times 3$ integer flux $N$ satisfying the F-term condition $(N\Omega)^T = N\Omega$. Thus, we have following background flux $F$,
\begin{align}
F = \pi [ N^T (Im \Omega)^{-1} ]_{ij} ( i dz^i \wedge d\bar{z}^j) = F_{z^i \bar{z}^j} (i dz^i \wedge d\bar{z}^j), 
\end{align}
where $F_{z^i \bar{z}^j} = F_{z^j \bar{z}^i}$.

We analyze the mass spectrum of 4D scalar modes resulting from the magnetized $T^6$ compactification. When $N Im \Omega > 0$, we have confirmed that the wave functions in eq.(\ref{eq: zero-mode_1+}) are eigenstates of the Laplacian on $T^6$. Their eigenvalue is equal to trace of $\frac{2\pi}{(2\pi R)^2}  [ N^T (Im \Omega)^{-1} ]$ which corresponds to the lowest energy, i.e. $\kappa_0^2$ in eq.(\ref{eq: Laplace_eigenfunction}). Hence, the lightest 4D scalar mode is given by their linear combination.   
We have shown the mass squared matrix of the 4D scalar modes in eq.(\ref{eq: 4D_scalar_mass_term}) where $\langle G \rangle_{z^i\bar{z}^{\bar{j}}}^{ab} = i F_{z^i \bar{z}^j}^{ab}(=iF_{z^i \bar{z}^j})$. We find that the lightest scalar mode corresponds to the eigenvector of the following mass squared matrix $\mathcal{M}^2$ with the smallest eigenvalue,
\begin{align}
\mathcal{M}^2 = \Delta +\frac{4\pi}{(2\pi R)^2}
\begin{bmatrix}
-A & O \\
O & A \\
\end{bmatrix}.
\end{align}
Here, $O$ is $3\times 3$ zero matrix and $A$ is $3 \times 3$ real-symmetric matrix defined as
\begin{align}
A = 
\begin{bmatrix}
[ N^T (Im \Omega)^{-1} ]_{11} & [ N^T (Im \Omega)^{-1} ]_{12} & [ N^T (Im \Omega)^{-1} ]_{13} \\
[ N^T (Im \Omega)^{-1} ]_{12} & [ N^T (Im \Omega)^{-1} ]_{22} & [ N^T (Im \Omega)^{-1} ]_{23} \\
[ N^T (Im \Omega)^{-1} ]_{13} & [ N^T (Im \Omega)^{-1} ]_{23} & [ N^T (Im \Omega)^{-1} ]_{33} \\
\end{bmatrix},
\end{align}
and $\Delta = \frac{2\pi}{(2\pi R)^2} \sum_{j = 1}^{3} [ N^T (Im \Omega)^{-1} ]_{jj}$.

We therefore obtain the D-term condition from eigenvalues of the above matrix.  
Since real symmetric matrices have real eigenvalues and are diagonalizable using orthogonal matrices, one can diagonalize mass matrix $\mathcal{M}^2$ as follows
\begin{align}
\mathcal{M}^2 
&= \frac{2\pi}{(2\pi R)^2}(\lambda_1 + \lambda_2 + \lambda_3) +\frac{4\pi}{(2\pi R)^2}
\begin{bmatrix}
-\lambda_1 & & & & & \\
 & -\lambda_2 & & & & \\
 & & -\lambda_3 & & & \\
 & & & \lambda_1 & & \\
 & & & & \lambda_2 & \\
 & & & & & \lambda_3 \\
\end{bmatrix}.
\end{align}
We note that $\lambda_i > 0, (i=1,2,3)$ is satisfied because of the positive definite condition $N^{T}({\rm Im\Omega})^{-1} > 0$.
D-term condition is that the smallest eigenvalue of $\mathcal{M}^2$ is zero. 
That is, if the largest eigenvalue $\lambda_i$ is equal to summation of the rest eigenvalues $\lambda_j$, $\lambda_k$ or $\lambda_i = \lambda_j + \lambda_k$, lightest mode is identified as the superpartner of the chiral fermion zero-modes. 
If the eigenvalue is negative corresponding to a tachyonic-mode, the system is unstable and there is no supersymmetry.

On the factorizable $T^6=T^2_1 \times T^2_2 \times T^2_3$, the D-term condition can be satisfied by tuning 
ratios of areas of $T^2_i$ \cite{Abe:2012ya,Abe:2012fj}.
However, we have no such free parameter in non-factorizable $T^6$.
That leads to severe constraints in models.

In the next subsection, we discuss three-generation models in magnetized $T^6/\mathbb{Z}_{7}$ and $T^6/\mathbb{Z}_{12}$ satisfying both F- and D-term conditions, hence phenomenologically attractive.

\subsection{4D $\mathcal{N}=1$ SUSY and magnetized $T^6/\mathbb{Z}_{7}$}
We perform numerical calculation to check whether there exist a model such that the D-term condition is satisfied. In the region $|n_{ii}| \leq 400$ and $|\det N| \leq 1.0 \times 10^{10}$, we find there are no such models except for the uninteresting case $N=O$, where $O$ denotes $3\times 3$ zero matrix. Therefore, magnetized $T^6/\mathbb{Z}_7$ model seems not suitable for realization of realistic models not only because it cannot reproduce three-generation, but also we do not find any D-flat models. 

\subsection{4D $\mathcal{N}=1$ SUSY and magnetized $T^6/\mathbb{Z}_{12}$}
We obtain three-generation models satisfying both F-term and D-term conditions.

When flux $N$ takes following values
\begin{align}
N = 
\begin{bmatrix}
-5 & 13 & 3 \\
13 & -5 & 3 \\
3 & 3 & 2 \\
\end{bmatrix},
\end{align}
the D-term condition is satisfied.
We find that $\det N = 36$, $ {\rm tr} \rho (G) = {\rm tr} \rho (G^2) = -\sqrt{3}i$, ${\rm tr} \rho (G^3) = 18$, ${\rm tr} \rho (G^4) = \sqrt{3} i$, ${\rm tr} \rho (G^6) = 36$ and the numbers of zero-modes are $[n_0, n_1, \cdots, n_{11}, n_{12}] = [9, 0, 2, 0, 9, 0, 3, 0, 9, 0, 4, 0]$.

Also when the flux is
\begin{align}
N = 
\begin{bmatrix}
-3 & 3 & 1 \\
 3 & -3 & 1 \\
 1 & 1 & -2 \\
\end{bmatrix},
\end{align}
the D-term condition is satisfied.
We find that $\det N = 12$, ${\rm tr} \rho (G) = {\rm tr} \rho (G^2) = -\sqrt{3}i$, ${\rm tr} \rho (G^3) = 6$, ${\rm tr} \rho (G^4) = \sqrt{3} i$, ${\rm tr} \rho (G^6) = 12$ and the numbers of zero-modes are $[n_0, n_1, \cdots, n_{11}, n_{12}] = [3, 0, 0, 0, 3, 0, 1, 0, 3, 0, 2, 0]$.

\section{Conclusion}
We have studied the zero-modes of non-factorizable $T^6/\mathbb{Z}_N$ orbifold models with background magnetic flux. We have classified zero-modes with $\mathbb{Z}_N$ charges in magnetized $T^6/\mathbb{Z}_N$ models by $Sp(6, \mathbb{Z})$ modular transformation.

We have focused on degenerated fermion zero-modes with the chirality $(+, +, +)$. To make it converge in magnetized D-brane model, we have used the facts that zero-modes are well-defined when $N Im\Omega$ is positive-definite under symmetric flux $N$.

Our results are important to check whether three-generation models in effective field theory exist or not systematically. 
We have constructed magnetized $T^6/\mathbb{Z}_N$ twisted orbifolds by generators of $Sp(6, \mathbb{Z})$ and symmetric flux $N$. By modular transformation of zero-modes and its representation, we can classify how many zero-modes with $\mathbb{Z}_7$ charges exist on each sectors. 

For $T^6/\mathbb{Z}_7$, we have not found any three-generation models when tr $\rho$ is equal to either one of $+1$, $-1$, or $-\sqrt{7}i$. This result can be proved by properties of number theory.  Therefore, if tr $\rho$ could take no other values than the above ones, it implies that magnetized $T^6/\mathbb{Z}_7$ model is not suitable for realizing three-generation models in four-dimensional effective field theory. In addition, we have never found any solutions when the models have no tachyonic-modes. 

For $T^6/\mathbb{Z}_{12}$, on the other hand, we have found some three-generation models. 
The F-term condition is satisfied in three-generation models when a determinant of flux $N$ is from $12$ to $48$. 
However, the D-term SUSY condition is satisfied  when $\det N$ is only $12$ or $36$ and $N$ are particular values. It means that there are three-generation models without tachyonic-modes.

In conclusion, we have seen that a few three-generation models can be realized on magnetized non-factorizable $T^6/\mathbb{Z}_N$ orbifolds. 
We have only used zero-modes with the chirality $(+, +, +)$, but there are zero-modes with other chirality such as $(+, -, -)$. 
Zero-modes with other chirality have different wave functions, and such analysis is non-trivial.
We need those wave functions to write Yukawa couplings in 4D low energy effective field theory.
Such studies on other chirality are beyond our scope, and we would study them elsewhere including 
realization of fermion masses and mixing angles.


\vspace{1.5 cm}
\noindent
{\large\bf Acknowledgement}\\

This work was supported by JSPS KAKENHI Grant Numbers JP22J10172 (SK), JP23K03375(TK) and JP20J20388 (HU), 
and 
JST SPRING Grant Number JPMJSP2119(KN and ST) .

\appendix

\section{$T^4/\mathbb{Z}_N$}

In this appendix, we study non-factorizable $T^4/\mathbb{Z}_N$ orbifolds.

\subsection{$\mathbb{Z}_N$ twists for $T^4/\mathbb{Z}_N$}

4D lattices have the modular symmetry, $Sp(4,\mathbb{Z})$.
Some of the Coxeter and generalized Coxeter elements 
can be realized by $Sp(4,\mathbb{Z})$ transformation.

Generators of $Sp(4, \mathbb{Z})$ are given by
\begin{align}
S
=
\begin{bmatrix}
O & I_2 \\
-I_2 & O \\
\end{bmatrix}
,\ 
T_i
=
\begin{bmatrix}
I_2 & B_i \\
O & I_2 \\
\end{bmatrix},\quad (i=1,2,3),
\end{align}
where $B_i$ are $2\times 2$ symmetric matrices defined by 
\begin{align}
B_1
=
\begin{bmatrix}
1 & 0 \\
0 & 0 \\
\end{bmatrix},
B_2
=
\begin{bmatrix}
0 & 0 \\
0 & 1 \\
\end{bmatrix},
B_3
=
\begin{bmatrix}
0 & 1 \\
1 & 0 \\
\end{bmatrix}.
\end{align}

Referring to Table \ref{tab:lattice}, we may expect to realize the $\mathbb{Z}_6, \mathbb{Z}_8$ and $\mathbb{Z}_{12}$ orbifolds with the $SO(8)$ Lie root lattice. However, we succeed in
describing only the $\mathbb{Z}_8$ in terms of $Sp(4,\mathbb{Z})$ as discussed in the following subsections.

\subsection{$T^4/\mathbb{Z}_8$ orbifold}
We consider the number of zero-modes on magnetized $T^4/\mathbb{Z}_8$ \cite{Kikuchi:2022psj} by following algebraic relation
\begin{align}
(ST_1 T^{-1}_2 T^{-1}_3)^8 = {\bf{1}_4}.
\end{align}
The invariant moduli $\Omega_8$  under the transformation $ST_1 T^{-1}_2 T^{-1}_3$ satisfy
\begin{align}
-(\Omega_8 + B_1 -B_2 -B_3)^{-1} = \Omega_8.
\end{align}
One of the solutions $\Omega_8 \in \mathcal{H}_2$ is given by
\begin{align}
\label{eq: moduli_SO(8)}
\Omega_8
=
\begin{bmatrix}
-\frac{1}{2} +\frac{i}{\sqrt{2}} & \frac{1}{2} \\
\frac{1}{2} & \frac{1}{2} +\frac{i}{\sqrt{2}}
\end{bmatrix}.
\end{align}
Flux $N$ is constrained by the F-term condition to the form,
\begin{align}
N=
\begin{bmatrix}
n_1 & \frac{n_2 -n_1}{2} \\
\frac{n_2 -n_1}{2} & n_2 \\
\end{bmatrix}
,\quad n_1 \equiv n_2 \pmod{2}.
\end{align}
 Then, $N$ is  $ST_1 T^{-1}_2 T^{-1}_3$ invariant. 
Furthermore, for the consistent transformation of zero-mode wave functions under $T_1T_2^{-1}T_3^{-1}$, additional constraints are imposed. That is, $N(B_1 - B_2 -B_3)$ is symmetric and its diagonal elements are all even. This leads to
\begin{equation}
    n_1 \equiv n_2 \equiv 0\quad  {\rm or}\quad 
    n_1 \equiv n_2 \equiv 2 \pmod{4}.
\end{equation}
Then, ${\rm det}N$ is always a multiple of 4. We obtain the following $ST_1 T^{-1}_2 T^{-1}_3$ transformation of zero-modes,
\begin{align}
\psi^{\vec{J}}_N (\Omega_8\vec{z} ,\Omega_8 ) 
=\frac{1}{\sqrt{\det N}} \sum_{\vec{K} \in \Lambda_N} e^{2\pi i\vec{J}^{T} N^{-1} \vec{K}} e^{\pi i \vec{K}^T N^{-1} B \vec{K}} \psi^{\vec{K}}_{N} (\vec{z}, \Omega_8),
\end{align}
where $B_1 -B_2 -B_3$ is denoted by $B$. We define $\Lambda_N$ as lattice spanned by $N\vec{e}_i$.
The representation of the algebraic structure $G=ST_1 T^{-1}_2 T^{-1}_3$ is given by
\begin{align}
\rho_{\vec{K_1} \vec{K_2}} (G) =
\frac{1}{\sqrt{\det N}}  e^{2\pi i\vec{K_1}^{T} N^{-1} \vec{K_2}} e^{\pi i \vec{K_1}^T N^{-1} B \vec{K_1}}.
\end{align}
Trace of the representation $\rho$ is
\begin{align}
{\rm tr} \rho 
=
\frac{1}{\sqrt{\det N}} \sum_{\vec{K} \in \Lambda_N} e^{2\pi i\vec{K}^{T} N^{-1} \vec{K}} e^{\pi i \vec{K}^T N^{-1} B \vec{K}}.
\end{align} 

Thus, we can obtain the zero-modes on magnetized $T^4/\mathbb{Z}_8$ that have $\mathbb{Z}_N$ charges from $\beta^n={\rm tr} \rho(G^n)$, $n=1,2,4$.
Table \ref{tb: T4/Z8} shows the zero-modes with $\det N$, ${\rm tr}\rho(G^n)$, the number of zero-modes in $\mathbb{Z}_N$ sector $n_k$. We see that there are three-generation models in the range of $16 \leq \det N \leq 32$.

\begin{table}[ht]
\centering
  \caption{The number of zero-modes on magnetized $T^4/\mathbb{Z}_{8}$}
  \label{tb: T4/Z8}
  \begin{tabular}{|l||c|r|c||r|c|r|c|r|c|r|c|}  \hline
    $D$  & $\beta^1$ & $\beta^2$ & $\beta^4$ & $n_0$ & $n_1$ & $n_2$ & $n_3$ & $n_4$ & $n_5$ & $n_6$ & $n_7$ \\ \hline 
   $4$& 0 & 0 & 4 & 1 & 0 & 1 & 0 & 1 & 0 & 1 & 0  \\ \hline 
   $8$& $\sqrt{2}i$ & 2 & 4 & 2 & 1 & 1 & 1 & 2 & 0 & 1 & 0   \\ \hline
   $16$& $\sqrt{2}i$ & 2 & 4 & \fbox{3} & 2 & 2 & 2 & \fbox{3} & 1 & 2 & 1 \\ \hline
   $28$& 0 & 0 & 4 & 4 & \fbox{3} & 4 & \fbox{3} & 4 & \fbox{3} & 4 & \fbox{3}  \\ \hline
   $32$& $\sqrt{2}i$ & 2 & 4 & 5 & 4 & 4 & 4 & 5 & \fbox{3} & 4 & \fbox{3} \\ \hline
   $36$& 0 & 0 & 4 & 5 & 4 & 5 & 4 & 5 & 4 & 5 & 4\\ \hline
  \end{tabular}
\end{table}

\subsection{$T^4/\mathbb{Z}_{12}$ orbifold}
The generalized Coxeter element $\mathbb{Z}_{12}$ on $SO(8)$ has the negative determinant.
That is, such an element is not included in $Sp(4,\mathbb{Z})$. 
Any non-vanishing magnetic fluxes are not consistent with the $\mathbb{Z}_{12}$ twist.
Thus, one can not describe the $\mathbb{Z}_{12}$ twist with our approach using $Sp(4,\mathbb{Z})$ transformation.

\subsection{$T^4/\mathbb{Z}_6$ orbifold}
Here we consider the $T^4/\mathbb{Z}_6$ orbifold defined by the Coxeter element $\mathbb{Z}_6$ on $SO(8)$. We have not succeeded in finding the corresponding $Sp(4,\mathbb{Z})$ generator. We discuss possible reasons behind this.

First, note that classifications of the fixed points of $Sp(4,\mathbb{Z})$ was studied in Ref.\cite{Gottschling:1961fp}. There are six independent zero-dimensional fixed points $\Omega_f \in \mathcal{H}_2$ being invariant under the actions of certain subgroups (stabilizer) of $Sp(4,\mathbb{Z})$. 

One of the fixed points is given by
\begin{equation}
\label{eq: S4_fixed_point}
    \Omega_f = 
    \begin{bmatrix}
     \eta & \frac{1}{2}(\eta - 1) \\
     \frac{1}{2}(\eta - 1) & \eta 
    \end{bmatrix} \in \mathcal{H}_2,
\end{equation}
where $\eta = \frac{1}{3}(1+i2\sqrt{2})$. Corresponding stabilizer group is generated by \cite{Ding:2021, Nilles:2021glx},
\begin{equation}
\label{eq: GL(2,3)_generators}
h_1 = 
    \begin{bmatrix}
     0 & 1 & 0 & 0 \\
     1 & 0 & 0 & 0 \\
     0 & 0 & 0 & 1 \\
     0 & 0 & 1 & 0
    \end{bmatrix},\quad  
h_2 = 
  \begin{bmatrix}
  -1 & 1 & 1 & 0 \\
  1 & 0 & 0 & 1 \\
  -1 & 0 & 0 & 0 \\
  1 & -1 & 0 & 1
  \end{bmatrix}.
\end{equation}
They form a finite group of order $48$ known as $GL(2,3)$.
One can find that $\Omega_f$ in eq.(\ref{eq: S4_fixed_point}) is equivalent to $\Omega_8$ in eq.(\ref{eq: moduli_SO(8)}) corresponding to the $SO(8)$ lattice. They are related by the following $Sp(4,\mathbb{Z})$ transformation,
\begin{equation}
\gamma \Omega_8 = \Omega_f,\quad 
\gamma = 
\begin{bmatrix}
1 &  0 & 0 & 0 \\
0 & -1 & 0 & 0 \\
1 & 0 & 1 & 0 \\
0 & 0 & 0 & -1
\end{bmatrix}\in Sp(4,\mathbb{Z}).
\end{equation}
This shows that $\Omega_f$ also corresponds to the same $SO(8)$ lattice, but with different basis choice related by $Sp(4,\mathbb{Z})$.

Then the stabilizer of $\Omega_8$ is also $GL(2,3)$ and its representation matrices are given by the matrix conjugation, $\gamma^{-1} h_i \gamma,\ (i=1,2)$.
The $\mathbb{Z}_6$ generator of the Coxeter element should be included in this stabilizer group if it is describable by $Sp(4,\mathbb{Z})$ generators. However, by the following argument we conclude that there is no such element.

One can compute the values of trace of all $48$ elements in $GL(2,3)$ generated by $h_i, (i=1,2)$ in eq.(\ref{eq: GL(2,3)_generators}). 
One obtains only even numbers.

On the other hand, the $\mathbb{Z}_6$ Coxeter element of $SO(8)$ is given by \cite{Kobayashi:1991rp}, 
\begin{equation}
\label{eq: Z_6_Coxeter}
\theta_{\mathbb{Z}_6} = 
    \begin{bmatrix}
    0 & 1 & 0 & 0 \\
    1 & 1 & 1 & 1 \\
    -1 & -1 & -1 & 0 \\
    -1 & -1 & 0 & -1
    \end{bmatrix},
\end{equation}
and its trace is odd, ${\rm tr}\ \theta_{\mathbb{Z}_6} = -1$. Note that the above matrix representation eq.(\ref{eq: Z_6_Coxeter}) assumes different basis choice compared with eq.(\ref{eq: S4_fixed_point}) although they span the same $SO(8)$ Lie lattice. If $\theta_{\mathbb{Z}_6}$ has an expression in terms of $Sp(4,\mathbb{Z})$, there exists $L \in GL(4,\mathbb{Z})$ such that 
\begin{equation}
    \theta_{\mathbb{Z}_6} \rightarrow \theta_{\mathbb{Z}_6}' = L^{-1} \theta_{\mathbb{Z}_6} L, \quad L \in GL(4,\mathbb{Z}),
\end{equation}
which corresponds to possible change of basis vectors. Since the trace is independent of the basis choice, we suspect that the $\mathbb{Z}_6$ Coxeter element of $SO(8)$ cannot be realized by $Sp(4,\mathbb{Z})$ transformation.

\section{Modular transformation}
\label{Modular}
We consider modular transformation $\gamma \in Sp(6,\mathbb{Z})$. The modular transformation is defined as the basis transformation of the lattice $\Lambda$ defining $T^6 \simeq \mathbb{C}^3/\Lambda$. The symplectic modular group $Sp(6, \mathbb{Z})$ is composed of $6\times 6$ integer matrices $\gamma$,
\begin{align}
\gamma
=
\begin{bmatrix}
A & B \\
C & D \\
\end{bmatrix}
\in Sp(6,\mathbb{Z}),
\end{align}
satisfying
\begin{equation}
    \gamma J \gamma^T = J,\quad 
    J
=
\begin{bmatrix}
0 & {\bf{1}}_3 \\
-{\bf{1}}_3 & 0 \\
\end{bmatrix}.
\end{equation}
We introduce the modular transformation $\gamma$ for the complex coordinates $\vec{z}$ and the complex structure moduli $\Omega$ under $\gamma$ as follows
\begin{align}
\begin{aligned}
\Omega &\rightarrow (A\Omega +B)(C\Omega +D)^{-1}, \\
\vec{z} &\rightarrow {(C\Omega +D)^{-1}}^{T}\vec{z},
\end{aligned}
\end{align}
where $A, B, C$, and $D$ are $3\times 3$ integer matrices and the generators $S$, $T_i$, $(i=1,2,\cdots,5,6)$ are given by
\begin{align}
S 
=
\begin{bmatrix}
O & {\bf{1}}_{3} \\
-{\bf{1}}_{3} & O \\
\end{bmatrix} , 
T_{i} =
\begin{bmatrix}
{\bf{1}}_3 & B_{i} \\
O & {\bf{1}}_3 \\
\end{bmatrix}.
\end{align}
The symmetric matrices $B_i$ are given by
\begin{align}
\begin{aligned}
&B_1 = 
\begin{bmatrix}
1 & 0 & 0 \\
0 & 0 & 0 \\
0 & 0 & 0 \\
\end{bmatrix}
, \quad 
B_2 = 
\begin{bmatrix}
0 & 0 & 0 \\
0 & 1 & 0 \\
0 & 0 & 0 \\
\end{bmatrix}
,\quad 
B_3 = 
\begin{bmatrix}
0 & 0 & 0 \\
0 & 0 & 0 \\
0 & 0 & 1 \\
\end{bmatrix}
,
\\
&B_4 = 
\begin{bmatrix}
0 & 1 & 0 \\
1 & 0 & 0 \\
0 & 0 & 0 \\
\end{bmatrix}
, \quad 
B_5 = 
\begin{bmatrix}
0 & 0 & 0 \\
0 & 0 & 1\\
0 & 1 & 0 \\
\end{bmatrix}
,\quad 
B_6 = 
\begin{bmatrix}
0 & 0 & 1 \\
0 & 0 & 0 \\
1 & 0 & 0 \\
\end{bmatrix}.
\end{aligned}
\end{align}
We will consider modular $S$ and $T_i$ transformations of zero-modes on magnetized $T^6$.

\subsection{The $S$ transformation}
Under the $S$ transformation, $\vec{z}=\vec{x}+\Omega \vec{y}$ and $\Omega$ behave as
\begin{align}
S : (\vec{z}, \Omega) \rightarrow
(\vec{z}_S, \Omega_S) = (-\Omega^{-1} \vec{z}, -\Omega^{-1}). 
\end{align}
We see that the complex coordinates, the moduli after the $S$ transformation are given by $\vec{z}_S = -\Omega^{-1} \vec{z} = \vec{x}_S -\Omega^{-1} \vec{y}_S$ and $\Omega_S = -\Omega^{-1}$. Also we obtain transformation of real coordinates $\vec{x}_S, \vec{y}_S$ as
\begin{align}
\begin{aligned}
\vec{x}_S &= -\vec{y}, \\
\vec{y}_S &= \vec{x}.
\end{aligned}
\end{align}

\subsubsection{Magnetic flux and F-term condition in $S$ transformation}
Magnetic flux on $T^6$ is defined by
\begin{align}
F
&= \frac{1}{2} p_{IJ}dX^{I} \wedge dX^{J} \notag \\
&= \frac{1}{2} (p_{xx})_{ij} dx^{i} \wedge dx^{j} + \frac{1}{2} (p_{yy})_{ij} dy^{i} \wedge dy^{j} + (p_{xy})_{ij} dx^{i} \wedge dy^{j},
\end{align}
where $X^{I} = (x^{i} ,y^{i})$, $i = 1, 2, 3$ is the real coordinate along the lattice.
Therefore the magnetic flux $F$ after the $S$ transformation is written by $\vec{x}_S = -\vec{y}$ and $\vec{y}_S = \vec{x}$
\begin{align}
F
&= \frac{1}{2} (p_{xx}^S)_{ij} dx^{i}_S \wedge dx^{j}_S + \frac{1}{2} (p_{yy}^S)_{ij} dy^{i}_S \wedge dy^{j}_S + (p_{xy}^S)_{ij} dx^{i}_S \wedge dy^{j}_S \notag \\
&= \frac{1}{2} (p_{xx}^S)_{ij} dy^{i} \wedge dy^{j} + \frac{1}{2} (p_{yy}^S)_{ij} dx^{i} \wedge dx^{j} + (p_{xy}^S)_{ij}^T dx^{i} \wedge dy^{j}.
\end{align}
We therefore find that 
\begin{align}
\begin{aligned}
p_{xx}^S &= p_{yy}, \\
p_{yy}^S &= p_{xx}, \\ 
p_{xy}^S &= (p_{xy})^T.
\end{aligned}
\end{align}
When we impose the condition $p_{xx} = p_{yy} = 0$, it is clear that magnetic flux is consistent with the $S$ transformation. Thus the flux $N = \frac{p_{xy}^T}{2\pi}$ is transformed under $S$ as
\begin{align}
S: N \rightarrow N^T
\end{align}
and F-term condition $(N\Omega)^T = N\Omega$ in the $S$ transformation is given by
\begin{align}
(N_S \Omega_S)^T  \notag 
&= (N^T (-\Omega^{-1}))^T \\ \notag
&= ((\Omega^{T})^{-1} N\Omega (-\Omega^{-1}))^T \notag \\
&= N_S \Omega_S,
\end{align}
where we use $(N\Omega)^T = N\Omega$ and the condition that $\Omega$ is symmetric.
Therefore we find that F-term condition as well as magnetic flux is consistent with the $S$ transformation.

\subsubsection{The S transformation of zero-modes} 
Zero-mode with the chirality $(+, +, +)$ is introduced by the Riemann-theta function with characteristics
\begin{align}
\psi^{\vec{J}, N}(\vec{z}, \Omega)
&= N \cdot e^{i\pi (N \vec{z})^T (Im\Omega)^{-1} \cdot Im(\vec{z})} \cdot
\theta 
\begin{bmatrix}
\vec{J}N^{-1} \\
0 \\
\end{bmatrix}
(N\vec{z}, N\Omega).
\end{align}
Since fluxes $N$ in our magnetized orbifold models are constrained to be symmetric, we consider $S$ transformation under the condition $N = N^T$. 

To come to the point, the $S$ transformation of zero-modes $\psi^{\vec{J}, N}(\vec{z}, \Omega)$ is written as
\begin{align}
\label{eq: zero-modes_S}
\psi^{\vec{J}}_{N} (-\Omega^{-1} \vec{z}, -\Omega^{-1})
&=\sqrt{\det (N^{-1}\Omega /i)} \sum_{\vec{K} \in \Lambda_{N}}e^{2\pi i\vec{J}^{T}N^{-1} \vec{K}} \psi^{\vec{K}}_{N} (\vec{z} ,\Omega) \notag \\
&= \frac{\sqrt{\det (-i\Omega)}}{\sqrt{\det N}} \sum_{\vec{K} \in \Lambda_{N}}e^{2\pi i\vec{J}^{T}N^{-1} \vec{K}} \psi^{\vec{K}}_{N} (\vec{z} ,\Omega),
\end{align}
where $\Lambda_N$ is a lattice spanned by $N\vec{e}_i$ and $\vec{e}_i$ are unit vectors.

In the following, we present a derivation of eq.(\ref{eq: zero-modes_S}).
We denoted the definitions of the Riemann-theta function. From the literature in mathematics \cite{Mumford:1983}, it is known that
\begin{align}
 \theta
(-\Omega^{-1} \vec{z} , -\Omega^{-1})
=\sqrt{\det(\Omega/i)}e^{\pi i \vec{z}^T \Omega^{-1} \vec{z}} \cdot \theta
(\vec{z}, \Omega),  \label{24}
\end{align}
holds where $\vec{z} \in \mathbb{C}^3, \Omega \in \mathcal{H}_3$. Note that we must take a branch of the square root which returns a positive number if $\Omega$ is purely imaginary.

When we replace $\vec{z}$ with $\vec{z} +N^{-1} \vec{J}$ in eq.\eqref{24}, we obtain
\begin{align}
\theta 
\begin{bmatrix}  
\vec{J}^{T} N^{-1} \\
\vec{0} \\
\end{bmatrix}
(-\Omega^{-1} \vec{z}, -\Omega^{-1}) 
= \sqrt{\det(\Omega /i)} \cdot e^{i\pi \vec{z}^{T} \Omega^{-1} \vec{z}} \cdot
\theta 
\begin{bmatrix}  
\vec{0} \\
\vec{J}^{T} N^{-1} \\
\end{bmatrix}
(\vec{z}, \Omega). \label{27}
\end{align}
Following relations were used
\begin{align}
\begin{aligned}
\label{eq: theta_characteristics_formula}
\theta
\begin{bmatrix}
\vec{a} \\
\vec{b}+\vec{b}^{\prime} \\
\end{bmatrix}
(\vec{z} , \Omega)
&= \theta
\begin{bmatrix}
\vec{a} \\
\vec{b} \\
\end{bmatrix}
(\vec{z} + \vec{b}^{\prime} , \Omega), \\
\theta
\begin{bmatrix}
\vec{a}+\vec{a}^{\prime}  \\
\vec{b}\\
\end{bmatrix}
(\vec{z} , \Omega)
&= e^{i\pi \vec{a}^{\prime T} \Omega \vec{a}^{\prime} + 2\pi i \vec{a}^{\prime T} (\vec{z} + \vec{b})} 
 \theta
\begin{bmatrix}
\vec{a} \\
\vec{b} \\
\end{bmatrix}
(\vec{z} + \Omega \vec{a}^{\prime} , \Omega),
\end{aligned}
\end{align}
where $\vec{a}^{\prime} ,\vec{b}^{\prime} \in \mathbb{R}^{3}$.

Now we replace $\Omega \in \mathcal{H}_{3}$ by $N^{-1} \Omega_I$ where $N$ is to be identified as the flux in magnetized D-brane models. Even if $\Omega_I$ is not an element of Siegel upper-half plane $\mathcal{H}_{3}$, our replacement is consistent if $N^{-1} \Omega_I \in \mathcal{H}_3$. In the following, we denote $\Omega_{I}$ by $\Omega$. Thus, we have
\begin{align}
&\theta 
\begin{bmatrix}  
\vec{J}^{T} N^{-1} \\
\vec{0} \\
\end{bmatrix}
(-\Omega^{-1} N \vec{z}, -\Omega^{-1} N) \notag \\
&= \sqrt{\det(N^{-1}\Omega /i)} \cdot e^{i\pi \vec{z}^{T} (N^{-1}\Omega)^{-1} \vec{z}} \cdot
\theta 
\begin{bmatrix}  
\vec{0} \\
\vec{J}^{T} N^{-1} \\
\end{bmatrix}
(\vec{z}, N^{-1}\Omega). \label{28}
\end{align}
The right-hand side of the eq.(\ref{28}) can be rewritten as 
\begin{dmath}
\theta 
\begin{bmatrix}  
\vec{0} \\
\vec{J}^{T} N^{-1} \\
\end{bmatrix}
(\vec{z}, N^{-1}\Omega)
=\sum_{\vec{\ell} \in \mathbb{Z}^3} e^{\pi i {\vec{\ell}}^T N^{-1} \Omega \vec{\ell}} e^{2\pi i {\vec{\ell}}^T (\vec{z} + N^{-1} \vec{J})}
=\sum_{\vec{K} \in \Lambda_N} \sum_{\vec{a} \in \mathbb{Z}^{3}} e^{\pi i {(N \vec{a} +\vec{K})}^T N^{-1} \Omega (N \vec{a} +\vec{K})} e^{2\pi i {(N \vec{a} +\vec{K})}^T (\vec{z} +N^{-1} \vec{J})}
=\sum_{\vec{K} \in \Lambda_N} e^{2\pi i\vec{K}^{T} N^{-1} \vec{J}} \sum_{\vec{a} \in \mathbb{Z}^{3}} e^{\pi i {(N \vec{a} +\vec{K})}^T N^{-1} \Omega (N \vec{a} +\vec{K})} e^{2\pi i {(N \vec{a} +\vec{K})}^T \vec{z}}
=\sum_{\vec{K} \in \Lambda_N} e^{2\pi i\vec{J}^{T} N^{-1} \vec{K}} 
\theta
\begin{bmatrix}  
\vec{K}^{T} N^{-1} \\
\vec{0} \\
\end{bmatrix}
(N \vec{z}, N\Omega),
\end{dmath}
where the summation variable $\vec{l}$ is decomposed into two variables $\vec{a} \in \mathbb{Z}^3$ and $\vec{K} \in \mathbb{Z}^3$ as $\vec{l} = N \vec{a} + \vec{K}$. Note that $\vec{K}$ are integer points inside the lattice $\Lambda_N$.

Notice that since we have F-term SUSY condition $(N\Omega)^T = N\Omega$ as well as $N^T = N$, $\Omega^T = \Omega$, the flux $N$ and the moduli $\Omega$ commute $[N, \Omega]=0$. From the relation, $\Omega^{-1} [N,\Omega] \Omega^{-1} =0$, and $N$, $\Omega^{-1}$ are also commutative.

We therefore find that eq. \eqref{28} can be expressed as 
\begin{align}
\label{eq: theta_S_trans}
&\theta 
\begin{bmatrix}  
\vec{J}^{T} N^{-1} \\
\vec{0} \\
\end{bmatrix}
(N(-\Omega^{-1} \vec{z}), N(-\Omega^{-1})) \notag \\
&=\sqrt{\det(N^{-1}\Omega /i)} \cdot e^{i\pi \vec{z}^{T} (N^{-1}\Omega)^{-1} \vec{z}} \sum_{\vec{K} \in \Lambda_N} e^{2\pi i \vec{J}^{T} N^{-1} \vec{K}} 
\cdot 
\theta
\begin{bmatrix}  
\vec{K}^{T} N^{-1} \\
\vec{0} \\
\end{bmatrix}
(N \vec{z}, N\Omega).
\end{align}

Next, we focus on the transformation of the phase of zero-modes. We find
\begin{align}
S : e^{i\pi (N\vec{z})^T \cdot (Im\Omega)^{-1} Im(\vec{z})} 
\rightarrow
e^{\pi i (-N\Omega^{-1}\vec{z})^{T} (Im (-\Omega^{-1}))^{-1} Im (-\Omega^{-1} \vec{z})}.
\end{align}
When this is multiplied by the exponential factor $e^{i\pi \vec{z}^{T} (N^{-1}\Omega)^{-1} \vec{z}}$ in the right-hand side of eq.(\ref{eq: theta_S_trans}), we get 
\begin{equation}
    e^{\pi i (-N\Omega^{-1}\vec{z})^{T} (Im (-\Omega^{-1}))^{-1} Im (-\Omega^{-1} \vec{z})} e^{i\pi \vec{z}^{T} (N^{-1}\Omega)^{-1} \vec{z}} = e^{i \pi (N\vec{z})^{T} (Im\Omega)^{-1}{Im \vec{z}}}.
\end{equation}
Finally, we find the $S$ transformation of zero-modes,
\begin{align}
\psi^{\vec{J}}_{N} (-\Omega^{-1} \vec{z}, -\Omega^{-1})
&= \frac{\sqrt{\det (-i\Omega)}}{\sqrt{\det N}} \sum_{\vec{K} \in \Lambda_{N}}e^{2\pi i\vec{J}^{T}N^{-1} \vec{K}} \psi^{\vec{K}}_{N} (\vec{z} ,\Omega).
\end{align}

\subsection{The $T$ transformation of zero-modes}
\label{appendix: modular_T_zero}
Under $T_i$, $(i =1,2,\cdots ,5,6)$ transformation, complex coordinates $\vec{z}$ and complex structure moduli $\Omega$ are transformed as
\begin{align}
T: (\vec{z},\Omega) \rightarrow (\vec{z}_T, \Omega_T) = (\vec{z},\Omega + B_i).
\end{align}
Let $\vec{z}_{T}$ represent complex coordinates after $T$ transformation and we omit the index $i$ in this subsection.  We obtain transformation of real coordinates,
\begin{align}
&\vec{z} = \vec{x} +\Omega \vec{y}  \rightarrow \vec{z}_{T} = \vec{x}_{T} +(\Omega +B )\vec{y}_{T} = \vec{z} = \vec{x} +\Omega \vec{y} \notag \\
&\leftrightarrow 
 \vec{x}_{T} = \vec{x} -B \vec{y} ,\quad \vec{y}_{T} = \vec{y}.
\end{align}
Therefore, the magnetic flux $F$ after the $T$ transformation is 
\begin{dmath}
F
= \frac{1}{2} (p_{xx}^{(T)})_{ij} dx_{T}^{i} \wedge dx_{T}^{j} + \frac{1}{2} (p_{yy}^{(T)})_{ij} dy_{T}^{i} \wedge dy_{T}^{j} + (p_{xy}^{(T)})_{ij} dx_{T}^{i} \wedge dy_{T}^{j}
= \frac{1}{2} (p_{xx}^{(T)})_{ij} dx^{i} \wedge dx^{j} + \frac{1}{2} [p_{yy}^{(T)} +Bp_{xx}^{(T)}B-(Bp_{xy}^{(T)})+(Bp_{xy}^{(T)})^{T}]_{ij} dy^{i} \wedge dy^{j} +[(p_{xy}^{(T)})-(p_{xx}^{(T)} B)]_{ij} dx^{i} \wedge dy_{j}. \label{38}
\end{dmath}
From eq.\eqref{38}, we can see the following transformation of the components $p_{xx}, p_{yy}, p_{xy}$,
\begin{align}
\begin{aligned}
p_{xx}^{(T)} &= p_{xx}, \\
p_{yy}^{(T)} &= p_{yy}  +Bp_{xx} B+(Bp_{xy}-(Bp_{xy}^{(T)})^{T}),  \\
p_{xy}^{(T)}  &= p_{xy}-(p_{xx} B).
\end{aligned}
\end{align}

Thus, we find that the following constraint is required for the condition $p_{xx} =p_{yy} =0$ to be consistent with $T$ transformation,
\begin{align}
\label{eq: B_pxy}
(B p_{xy})^{T} = B p_{xy}.
\end{align}

Noting the Dirac's quantization condition $p_{xy} = 2\pi N^{T}$, and the fact that matrices $B$ in $Sp(6, \mathbb{Z})$ are symmetric, we can write eq.(\ref{eq: B_pxy}) as
\begin{align}
(NB)^{T} = NB. \label{41}
\end{align}
Then it follows that F-term SUSY condition is consistent with the $T$ transformation
\begin{dmath}
(N_T \Omega_{T})^{T} 
= (\Omega +B)^{T} N^{T}
= (N\Omega)^{T} +(NB)^{T}
=N\Omega +NB
=N_{T} \Omega_{T}.
\end{dmath}

Next we consider the $T$ transformation of zero-modes. In the following, we require that  all diagonal components of $NB$ are even, and then we obtain 
\begin{align}
\begin{aligned}
&\theta(N\vec{z}, N(\Omega +B))  \\
&= \sum_{\vec{m} \in \mathbb{Z}^{3}} \exp(\pi i \vec{m}^{T} 
N(\Omega +B) \vec{m} +2\pi i \vec{m}^{T} N\vec{z}) \\
&=\sum_{\vec{m} \in \mathbb{Z}^{3}} e^{\pi i \vec{m}^{T} 
N\Omega \vec{m} +2\pi i \vec{m}^{T} N\vec{z}} e^{\pi i\vec{m}^{T} NB \vec{m}}  \\
&= \sum_{\vec{m} \in \mathbb{Z}^{3}} e^{\pi i \vec{m}^{T} 
N\Omega \vec{m} +2\pi i \vec{m}^{T} N\vec{z}} \cdot 1  \\
&= \theta(N\vec{z}, N\Omega).
\end{aligned}
\end{align}
We replace complex coordinates $\vec{z}$ with $\vec{z} +(\Omega +B)N^{-1 T} \vec{J}$,
\begin{dmath}
\theta(N(\vec{z} +(\Omega +B)N^{-1 T} \vec{J}), N(\Omega +B))
=\theta(N(\vec{z} +(\Omega +B)N^{-1 T} \vec{J}), N\Omega ).
\end{dmath}
By use of eq.(\ref{eq: theta_characteristics_formula}), we can express it as
\begin{dmath}
\theta
\begin{bmatrix}
\vec{J}^{T} N^{-1} \\
\vec{0} \\
\end{bmatrix}
(N\vec{z}, N(\Omega +B))
= e^{-\pi i\vec{J}^{T} N^{-1} B\vec{j}} 
\theta
\begin{bmatrix}
\vec{J}^{T} N^{-1} \\
\vec{0} \\
\end{bmatrix}
(N\vec{z} +B\vec{J} , N\Omega)
=e^{-\pi i\vec{J}^{T} N^{-1} B\vec{j}} 
\theta
\begin{bmatrix}
\vec{J}^{T} N^{-1} \\
\vec{J}^{T} B \\
\end{bmatrix}
(N\vec{z} , N\Omega)
=e^{\pi i\vec{J}^{T} N^{-1} B\vec{J}} 
\theta
\begin{bmatrix}
\vec{J}^{T} N^{-1} \\
\vec{0} \\
\end{bmatrix}
(N\vec{z} , N\Omega).
\end{dmath}
The phase factor $e^{\pi i (N\vec{z})^{T} (Im \Omega)^{-1} Im\vec{z}}$ of the zero-modes is clearly invariant under the $T$ transformation. Therefore, we find the following $T$ transformation of zero-modes 
\begin{align}
\psi^{\vec{J}}_{N} (\vec{z} ,\Omega +B) = e^{\pi i\vec{J}^{T} N^{-1} B \vec{J}} \psi^{\vec{J}}_{N} (\vec{z} , \Omega).
\end{align}

\end{document}